\pdfoutput=1

\documentclass[aps,prd,twocolumn,10pt,nofootinbib,longbibliography]{revtex4-1}  

\usepackage[utf8x]{inputenc}
\usepackage[T1]{fontenc}

\usepackage[english]{babel}

\usepackage{float}
\usepackage{graphicx}   			
\usepackage{subfigure} 				
\usepackage{dcolumn}   				
\usepackage{amssymb}   				
\usepackage{amsmath}   				
\usepackage{bm}        				
\usepackage{mathtools} 				
\usepackage{mathrsfs}				
\usepackage[hidelinks]{hyperref}  	
\usepackage{mathptmx}				
\usepackage{booktabs}				
\usepackage{xcolor}					

\newcommand{\D}{\mathrm{d}}				
\hypersetup{							
    colorlinks,
    linkcolor={blue!50!black},
    citecolor={blue!50!black},
    urlcolor={blue!50!black}
}

\let\vec\mathbf							
\renewcommand\arraystretch{1.5}
\addto\extrasenglish{%

}

\begin{document}



\title{Exotic compact object behavior in black hole analogues}

\author{Carlos A. R. Herdeiro}
\author{Nuno M. Santos}
\affiliation{Centro de Astrof\'{i}sica e Gravitaç\~{a}o $-$ CENTRA, Departamento de F\'{i}sica, Instituto Superior T\'{e}cnico $-$ IST, Universidade de Lisboa $-$ UL, Avenida Rovisco Pais 1, 1049, Lisboa, Portugal
}%

\begin{abstract}
Classical phenomenological aspects of acoustic perturbations on a draining bathtub geometry where a surface with reflectivity $\mathcal{R}$ is set at a small distance from the would-be acoustic horizon, which is excised, are addressed. Like most exotic compact objects featuring an ergoregion but not a horizon, this model is prone to instabilities when $|\mathcal{R}|^2\approx 1$. However, stability can be attained for sufficiently slow drains when $|\mathcal{R}|^2\lesssim70\%$. It is shown that the superradiant scattering of acoustic waves is more effective when their frequency approaches one of the system's quasi-normal mode frequencies. 
\end{abstract}

\date{February 2019}

\maketitle

\section{\label{sec:level1}Introduction}

Analogue models for gravity have proven to be a powerful tool in understanding and probing several classical and quantum phenomena in curved spacetime, namely the emission of Hawking radiation and the amplification of bosonic field perturbations scattered off spinning objects, commonly dubbed superradiance. It was Unruh who first drew an analogue model for gravity relating the propagation of sound waves in fluid flows with the kinematics of waves in a classical gravitational field \cite{Unruh:1981}. This seminal proposal unfolded a completely unexpected way of exploring gravity and opened a track to test it in the laboratory. A summary of the history and motivation behind analogue gravity can be found in \cite{AnGravLivingR:2005}. 

The analogy between the propagation of sound waves in a non-relativistic, irrotational, inviscid, barotropic fluid and the propagation of a minimally coupled massless scalar field in a curved Lorentzian geometry was primarily established by Visser \cite{Visser:1998,Visser:1999}, who also formulated the concepts of acoustic horizon, ergoregion and surface gravity in analogue models. The correspondence is purely kinematic, i.e. only kinematic aspects of general relativity, such as event horizons, apparent horizons and ergoregions, are carried over into fluid mechanics. The effective geometry of the flow is mathematically encoded in a Lorentzian metric, commonly dubbed \textit{acoustic metric}, governed by the fluid equations of motion and not by Einstein's field equations. In other words, the dynamic aspects of general relativity do not map into fluid mechanics. This partial isomorphism thus offers a rather simple way to disentangle the kinematic and dynamic contributions to some important phenomena in general relativity.

Dumb holes, the acoustic analogues for black holes (BHs), bear several structural and phenomenological similarities to BHs. For instance, the draining bathtub vortex \cite{Visser:1998,SchutzholdUnruh:2002}, or simply draining bathtub, which features both an ergoregion and an acoustic horizon, is a Kerr BH analogue. The model describes the two-dimensional flow of a non-relativistic, locally irrotational, inviscid, barotropic fluid swirling around a drain. The fluid velocity increases monotonically downstream. The region where the magnitude of the fluid velocity exceeds the speed of sound defines the ergoregion. Nearer the drain is the acoustic horizon, which comprises the set of points in which the radial component of the fluid velocity equals the speed of sound. Any sound wave produced inside the acoustic horizon cannot escape from the region around the drain. 

The phenomenology of the draining bathtub model has been widely addressed over the last two decades. For instance, works on quasi-normal modes (QNMs) \cite{CardosoLemosYoshida:2004,CrispinoDolanOliveira:2012}, absorption processes \cite{CrispinoDolanOliveira:2010} and superradiance \cite{BasakMajumdar:2003,Basak:2003,BertiCardosoLemos:2004,Richartz:2013} showed that this vortex geometry shares many properties with Kerr spacetime.

Kerr BHs are stable against linear bosonic perturbations \cite{TeukolskyIII:1974,Whiting:1989,BertiCardoso:2009,Teukolsky:2015}. The event horizon absorbs any negative-energy physical states which may form inside the ergoregion and would otherwise trigger an instability. In fact, as first shown by Friedman \cite{Friedman:1978}, asymptotically-flat, stationary solutions to Einstein's field equations possessing an ergoregion but not an event horizon may develop instabilities, usually called ergoregion instabilities, when linearly interacting with scalar and electromagnetic field perturbations, especially if rapidly spinning. Ergoregion instabilities are known to affect a plethora of exotic compact objects (ECOs) \cite{Vilenkin:1978,CominsSchutzChandra:1978,EriguchiYoshida:1996,AnderssonRuoffKokkotas:2004,CardosoPaniCadoniCavaglia:2008,CardosoPaniCadoniCavaglia:2008,CardosoPaniCadoniCavagliaII:2008,BarausseBertiCardosoPani:2010,MaggioPaniFerrari:2017,Maggio:2018}. These are loosely defined in the literature as objects without event horizon, more massive than neutron stars and suffciently dim not to have been observed by state-of-the-art electromagnetic telescopes and detectors yet. Examples include boson stars \cite{Kaup:1968}, anisotropic stars \cite{BowersLiang:1974}, wormholes \cite{Thorne:1988}, gravastars \cite{Mazur:2004}, fuzzballs \cite{Mathur:2005}, black stars \cite{Visser:2008}, superspinars \cite{Gimon:2009}, Proca stars \cite{BritoCardosoHerdeiroRadu:2016}, collapsed polymers \cite{Brustein:2017}, $2-2$ holes \cite{HoldomRen:2017} and AdS bubbles \cite{Danielsson:2017}. All these ECOs may be prone to light ring instabilities \cite{BertiCunhaHerdeiro:2017}. The timescale of such instability, however, is unknown. An exhaustive review including references (if existing for a given ECO model) on the formation, stability and electromagnetic and gravitational signatures of such objects can be found in \cite{CardosoPani:2017}.

The recent gravitational-wave (GW) detections \cite{GWEvent1,GWEvent2,GWEvent3,GWEvent4,GWEvent5,GWCatalog} from compact binary coalescences heralded the dawn of a brand new field in astronomy and astrophysics. The newborn era of precision GW physics is expected to probe strong-field gravity spacetime regions in the vicinity of compact objects and, most importantly, to provide the strongest evidence of event horizons. While current electromagnetic-wave and GW observations do support the existence of BHs, some other exotic alternatives are not excluded yet $-$ not even those which do not feature an event horizon. There has been a renewed interest in these exotic alternatives over the last decades because some ECOs can mimic the physical behavior of BHs, namely those whose near-horizon geometry slightly differs from that of BHs, such as Kerr-like ECOs \cite{MaggioPaniFerrari:2017,Maggio:2018}. These objects feature an ergoregion and are endowed with a surface with reflective properties at a microscopic or Planck distance from the would-be event horizon of the corresponding Kerr BH. The presence of an ergoregion and the absence of an event horizon are the two key ingredients for ergoregion instabilities to develop. Indeed, perfectly-reflecting Kerr-like ECOs exhibit some exponentially-growing QNMs and are unstable against linear perturbations in some region of parameter space.  However, the unstable modes can be mitigated or neutralized when the surface is not perfectly- but partially-reflecting, i.e. when dissipative effects are considered.

This work focuses on classical phenomenological aspects of a draining bathtub model whose acoustic horizon is replaced by a surface with reflective properties. The physical behavior of this «toy model» is quite similar to that of Kerr-like ECOs. Although the experimental realization of such system is not evident at present, it is still fruitful to explore this theoretical setup, as it may provide some insights into the generic features of ECOs and even be useful for possible future experiments. 

The present paper is organized as follows. \autoref{sec:level2} covers in brief the main mathematical and physical features of the draining bathtub geometry, introduces the equation governing acoustic perturbations and discusses the properties of its solutions. Numerical results regarding QNM frequencies and amplification factors of the system are presented \autoref{sec:level4}, complemented with a low-frequency analytical treatment in the Appendices \ref{sec:level3} and \ref{apx:2}.


\section{\label{sec:level2}Draining bathtub model}
\subsection{Acoustic metric}

The draining bathtub geometry introduced by Visser \cite{Visser:1998} describes the irrotational flow of a barotropic and inviscid fluid with background density $\rho_0$ in a plane with a sink at the origin. The irrotational nature of the flow together with the conservation of angular momentum require the background density to be constant, i.e. position-independent. As a result, the background pressure $p_0$ and the speed of sound $c$ are also constant throughout the flow. Furthermore, it follows from both the equation of continuity and the conservation of angular momentum that the unperturbed velocity profile $\vec{v}_0$ of the flowing fluid is given in polar coordinates $(r,\phi)$ by
\begin{equation}
\label{eq:velocity}
\vec{v}_0\equiv v_0^{(r)}~\vec{e}_r+v_0^{(\phi)}~\vec{e}_\phi,
\end{equation}
with $v_0^{(r)}=-A/r$ and $v_0^{(\phi)}=B/r$, where $A,B\in\mathbb{R}^+$. $\vec{e}_r$ and $\vec{e}_\phi$ are the radial and azimuthal unit vectors, respectively. A general velocity profile $\vec{v}$, allowing for perturbations, can be written as the gradient of a velocity potential $\Psi$, i.e. $\vec{v}=\nabla\Psi$. When $\vec{v}=\vec{v}_0$,  $\Psi=\Psi_0(r,\theta)\equiv A\log r+B\phi$ (apart from a constant of integration).

In polar coordinates $(t,r,\phi)$, the line element of the draining bathtub model has the form \cite{SchutzholdUnruh:2002,VisserWeinfurtner:2005}
\begin{align}
\label{eq:RB-metric}
\D{s^2}=-c^2\D t^2+\left(\D r+\frac{A}{r}\D t\right)^2+\left(r\D\phi-\frac{B}{r}\D t\right)^2,
\end{align}
where the constant prefactor $(\rho_0/c)^2$ has been omitted. Eq. \eqref{eq:RB-metric} has the Painlevé-Gullstrand form \cite{Berti:2005}. 

The flow is stationary and axisymmetric, i.e. the metric tensor in Eq. \eqref{eq:RB-metric} does not depend explicitly on $t$ nor on $\phi$. In polar coordinates, the Killing vectors associated with these continuous symmetries are $\bm{\xi}_t\equiv\partial_t$ and $\bm{\xi}_\phi\equiv\partial_\phi$, respectively. This geometry has an acoustic horizon located at $r=r_H\equiv A/c$ and a region of transonic flow defined by $r_H<r<r_E$, where $r_E=\sqrt{A^2+B^2}/c$ is the location of the ergosphere\footnote{The speed of the flowing fluid is $\phi-$independent and given by $v_0\equiv\|\vec{v}_0\|=\sqrt{A^2+B^2}/r$. It equals the speed of sound at $r=r_E$ and exceeds it everywhere in the region $r_H<r<r_E$.}.

Performing the coordinate transformation
\begin{align}
\label{eq:Polar-BL}
\D\tilde{t}&=\D t-\frac{Ar}{c^2r^2-A^2}\D r,\\
\D\tilde{\phi}&=\D\phi-\frac{AB}{r(c^2r^2-A^2)}\D r,
\end{align}
one can cast Eq. \eqref{eq:RB-metric} in the form \cite{Basak:2003,BertiCardosoLemos:2004,DolanOliveira:2013}
\begin{align}
\label{eq:RB-metric-BL}
\D{s^2}=-f(r)\D\tilde{t}^2+[g(r)]^{-1}\D r^2-\frac{2B}{c}\D\tilde{t}\D\tilde{\phi}+r^2\D\tilde{\phi}^2,
\end{align}
where
\begin{align}
f(r)=1-\frac{A^2+B^2}{c^2r^2}
\quad\text{and}\quad
g(r)=1-\frac{A^2}{c^2r^2}.
\end{align}
The coordinates $(\tilde{t},r,\tilde{\phi})$ are equivalent to the Boyer-Lindquist coordinates commonly used to write the Kerr metric.

\subsection{Acoustic perturbations}

\subsubsection{Model}

Perturbations to the steady flow can be encoded in the velocity potential by adding a term to it, i.e. by considering $\Psi=\Psi_0+\Phi$, where $\Phi$ is the perturbation. The equations of motion governing an acoustic perturbation in the velocity potential of an irrotational flow of a barotropic and inviscid fluid are the same as the Klein-Gordon equation for a minimally coupled massless scalar field propagating in a Lorentzian geometry \cite{Visser:1998}. In effect, the perturbation $\Phi$ in the velocity potential satisfies the equation
\begin{equation}
	\Box\Phi=\frac{1}{\sqrt{-g}}\partial_\mu\left(\sqrt{-g}\,g^{\mu\nu}\partial_\nu\Phi\right)=0,
	\label{EOM-massless-Phi}
\end{equation}
where $\Box\equiv\nabla_\mu\nabla^\mu$ is the D'Alambert operator and $\nabla_\mu$ denotes the covariant derivative. In the present case, the index $\mu$ runs from $0$ to $2$, with $0$ referring to the time coordinate $t$ and $1$ and $2$ to the spatial coordinates $r$ and $\phi$, respectively.

One is interested in $(2+1)-$dimensional objects whose geometry is described by Eq. \eqref{eq:RB-metric} from $r_0$ to infinity, where $r_0$ is the location of a surface with reflectivity $\mathcal{R}$. Such objects will hereafter be referred to as ECO-like vortices. Perfectly-reflecting and perfectly-absorbing surfaces are defined by $|\mathcal{R}|=1$ and $\mathcal{R}=0$, respectively. The model sets the surface at a small distance from the would-be acoustic horizon, i.e. at $r_0=r_H(1+\delta)$, where $0<\delta\ll 1$. The requirement that $\delta\ll 1$ allows ECO-like vortices to feature an ergoregion ($r_0<r_E$)\footnote{ECO-like vortices reduce to the draining bathtub when $\delta=0$ and $\mathcal{R}=0$.}. 

For the sake of simplicity, the uniform scalings $r\rightarrow Ar/c$ and $B\rightarrow B/A$ will hereafter be adopted \cite{CardosoLemosYoshida:2004,BertiCardosoLemos:2004}. Note that these linear transformations are equivalent to set $A=c=1$ in Eqs. \eqref{eq:RB-metric} and \eqref{eq:RB-metric-BL}. 

\subsubsection{Perturbation equation}

From the existence of the Killing vectors $\bm{\xi}_t$ and $\bm{\xi}_\phi$, one can separate the $t-$ and $\phi-$dependence of the field $\Phi$, which in turn can be expressed as a superposition of modes with different frequencies $\omega$ and periods in $\phi$, i.e.
\begin{align}
	\label{eq:Phi-Fourier}
	\Phi(t,r,\phi)=\sum_{m=-\infty}^{+\infty} \int_{-\infty}^{+\infty} \D\omega~e^{-i\omega t}R_{\omega m}(r)e^{+im\phi},
\end{align}
where $R_{\omega m}(r)$, dubbed radial function, is a function of the radial coordinate only and depends on $B$, $\omega$ and $m$. The radial function satisfies the ordinary differential equation (ODE) \cite{BasakMajumdar:2003}
\begin{equation}
\label{eq:RadODE-R}
\left[\frac{\D^2}{\D r^2}+\mathcal{K}_1(r)\frac{\D}{\D r}+\mathcal{K}_2(r)\right]R_{\omega m}(r)=0,
\end{equation}
where
\begin{align}
\mathcal{K}_1(r)&=\frac{1+r^2+2i(mB-\omega r^2)}{r(r^2-1)},\\
\mathcal{K}_2(r)&=\frac{\omega^2r^4-m^2r^2+m^2B^2-2mB\omega r^2-2imB}{r^2(r^2-1)}.
\end{align}
Defining a new radial function $S_{\omega m}(r)$ as
\begin{equation}
R_{\omega m}(r)=S_{\omega m}(r)e^{\frac{i}{2}\left[(\omega-mB)\log(r^2-1)+2mB\log(r)\right]},
\end{equation}
Eq. \eqref{eq:RadODE-R} becomes
\begin{equation}
\label{eq:RadODE-S}
\left[\frac{\D^2}{\D r^2}+\mathcal{L}_1(r)\frac{\D}{\D r}+\mathcal{L}_2(r)\right]S_{\omega m}(r)=0,
\end{equation}
where
\begin{align}
\mathcal{L}_1(r)&=\frac{r^2+1}{r(r^2-1)},\\
\mathcal{L}_2(r)&=\frac{\omega^2r^4-(m^2+2mB\omega)r^2+m^2(1+B^2)}{(r^2-1)^2}.
\end{align}

It is useful to introduce a tortoise coordinate defined by the condition \cite{BasakMajumdar:2003}
\begin{equation}
\frac{\D r_*}{\D r}=[g(r)]^{-1}=\left(1-\frac{1}{r^2}\right)^{-1}.
\end{equation}
Explicitly, the tortoise coordinate is given by
\begin{equation}
\label{eq:tortoise}
r_*(r)=r+\frac{1}{2}\log\left|\frac{r-1}{r+1}\right|,
\end{equation}
which maps the acoustic horizon at $r_H=1$ to $r_*\rightarrow-\infty$ and $r\rightarrow+\infty$ to $r_*\rightarrow+\infty$. Together with the definition \cite{BasakMajumdar:2003}
\begin{equation}
\label{eq:RadFun-SH}
S_{\omega m}(r)=\frac{H_{\omega m}(r)}{\sqrt{r}},
\end{equation}
Eq. \eqref{eq:RadODE-S} can be written as
\begin{equation}
\label{eq:RadODE-H}
\left[\frac{\D^2}{\D r_*^2}+\mathcal{V}(r)\right]H_{\omega m}(r)=0,
\end{equation}
where the effective potential $\mathcal{V}(r)$ is given by
\begin{equation}
\mathcal{V}(r)=\left(\omega-\frac{mB}{r^2}\right)^2-\frac{g(r)}{4r^2}\left(4m^2-1+\frac{5}{r^2}\right),
\end{equation}
which has the asymptotic behavior
\begin{align}
	\renewcommand\arraystretch{1}
	\mathcal{V}(r)\sim
	\left\{
	\begin{array}{ll}
        \omega^2,& r_*\rightarrow+\infty\\
        \varpi^2,& r_*\rightarrow-\infty
	\end{array}\right.,
	\label{QNM-system-pot-exp}
\end{align}
with $\varpi\equiv\omega-mB$. $B$ coincides with the angular velocity of the would-be acoustic horizon.

\subsubsection{Asymptotic solutions}

The presence of a surface with reflectivity $\mathcal{R}$ requires solutions to Eq. \eqref{eq:RadODE-H} to be a superposition of ingoing and outgoing waves at $r=r_0$. Thus, following the notation in \cite{Macedo:2018}, general solutions have the asymptotics
\begin{align}
	&H_{\omega m}(r)\sim
	\left\{
	\begin{array}{ll}
		e^{-i\varpi r_*}+\mathcal{R}e^{+i\varpi(r_*-2r_0^*)},& r_*\rightarrow r_0^*\\
        A_s^-e^{-i\omega r_*}+A_s^+e^{+i\omega r_*},& r_*\rightarrow+\infty
	\end{array}\right.,
	\label{QNM-system-sol-in-up}
\end{align}
where $r_0^*\equiv r_*(r_0)<0$. For the sake of simplicity, although $\mathcal{R}$ may depend on $\omega$ and/or $r_0$ \cite{CardosoWeinfurtner:2016,Macedo:2018,BenoneCrispinoHerdeiro:2018}, this work will only focus on constant-valued ($\omega-$ and $r_0-$independent) reflectivities.

Perfectly-reflecting ($|\mathcal{R}|^2=1$) boundary conditions (BCs), generically known as Robin BCs, are given by \cite{HerdeiroFerreira:2018}
\begin{align}
	\cos(\xi)\,H_{\omega m}(r_0)+\sin(\xi)\,H_{\omega m}'(r_0)=0,
	\label{QNM-RBC}
\end{align}
where $\xi\in[0,\pi)$ and the prime denotes differentiation with respect to $r$. Note that $\xi=0$ corresponds to a Dirichlet BC (DBC), i.e. $H_{\omega m}(r_0)=0$, whereas $\xi=\pi/2$ refers to a Neumann BC (NBC) imposed on $H_{\omega m}(r)$ at $r=r_0$, i.e. $H_{\omega m}'(r_0)=0$. Equivalently, DBCs (NBCs) can be defined by $\mathcal{R}=-1$ ($\mathcal{R}=1$)\footnote{Plugging $H_{\omega m}(r_0)=(1+\mathcal{R})e^{-i\varpi r_0^*}$ and $H_{\omega m}'(r_0)=-i\varpi(1-\mathcal{R})e^{-i\varpi r_0^*}$ into Eq. \eqref{QNM-RBC}, one can show that $\mathcal{R}=-[\cos(\xi)-i\varpi\sin(\xi)]/[\cos(\xi)+i\varpi\sin(\xi)]$.}. Perfectly-reflecting BCs will hereafter be specialized to DBCs and NBCs only.

The solution in Eq. \eqref{QNM-system-sol-in-up} may be written as a superposition of modes with asymptotics \cite{Vilenkin:1978}
\begin{align}
	&H_{\omega m}^+(r)\sim
	\left\{
	\begin{array}{ll}
		e^{-i\varpi r_*},& r_*\rightarrow r_0^*\\
        A_\infty^-e^{-i\omega r_*}+A_\infty^+e^{+i\omega r_*},& r_*\rightarrow+\infty
	\end{array}\right.
	\label{QNM-system-sol-in}\\
	&H_{\omega m}^-(r)\sim
	\left\{
	\begin{array}{ll}
		A_h^-e^{-i\varpi r_*}+A_h^+e^{+i\varpi r_*},& r_*\rightarrow r_0^*\\
        e^{+i\omega r_*},& r_*\rightarrow+\infty
	\end{array}\right.
	\label{QNM-system-sol-up}
\end{align}
$H_{\omega m}^+$ is a draining bathtub solution ($\mathcal{R}=0$) of the scattering problem, whereas $H_{\omega m}^-$ is an ECO solution of the QNM eigenvalue problem (since it is a superposition of ingoing and outgoing waves at the reflective surface).

From the constancy of the Wronskians of Eq. \eqref{eq:RadODE-H}, one can write the following useful relations between the coefficients $A_s^\pm$, $A_h^\pm$ and $A_\infty^\pm$:
\begin{align}
&\omega A_\infty^-=\varpi A_h^+,\label{eq:Wrskn1}\\
&\omega A_\infty^+=-\varpi A_h^{-*},\label{eq:Wrskn2}\\
&\omega A_s^-=\varpi(A_h^+-A_h^-\mathcal{R}e^{-2i\varpi r_0^*}),\label{eq:Wrskn3}\\
&\omega(A_s^+A_\infty^--A_s^-A_\infty^+)=\varpi\mathcal{R}e^{-2i\varpi r_0^*},\\
&\omega(|A_s^-|^2-|A_s^+|^2)=\varpi(1-|\mathcal{R}|^2).\label{eq:Wrskn5}
\end{align}

\subsubsection{Superradiance}

The amplification factor in a scattering process is defined by \cite{BritoCardosoPani:2015}
\begin{align}
\label{eq:AmpFactors}
Z(\omega,\mathcal{R})=\left|\frac{A_s^+}{A_s^-}\right|^2-1=-\left(1-\frac{mB}{\omega}\right)\frac{1-|\mathcal{R}|^2}{|A_s^-|^2},
\end{align}
where the last equality follows from the Wronskian relation in Eq. \eqref{eq:Wrskn5}. The amplitude of the reflected wave is greater than that of the incident wave at infinity (i.e. $Z>0$) when $0<\omega<mB$. Note that the superradiance condition does not depend on $\mathcal{R}$. 

\subsubsection{Quasi-normal modes}

Once physical BCs at $r_*\rightarrow r_0^*$ and $r_*\rightarrow+\infty$  are imposed, Eq. \eqref{eq:RadODE-H} defines an eigenvalue problem. If one requires purely outgoing waves at infinity,
\begin{equation}
	\label{eq:BC-infty}
    H_{\omega m}(r)\sim e^{+i\omega r_*},
    \quad
    r_*\rightarrow+\infty,
\end{equation}
the eigenvalues, the characteristic frequencies $\omega_\text{QNM}$, are called QNM frequencies and the corresponding perturbations $\Phi$ are dubbed QNMs \cite{BertiCardoso:2009}. The set of all eigenfrequencies is often referred to as QNM spectrum. The QNM frequencies $\omega_\text{QNM}$ are in general complex, i.e. $\omega_\text{QNM}=\omega_R+i\omega_I$, where $\omega_R\equiv\operatorname{Re}\{\omega_\text{QNM}\}$ and $\omega_I\equiv\operatorname{Im}\{\omega_\text{QNM}\}$. The sign of $\omega_I$ defines the stability of the QNM. According to the convention for the Fourier decomposition in Eq. \eqref{eq:Phi-Fourier}, if: $\omega_I<0$, the mode is stable and $\tau_\text{dam}\equiv 1/|\omega_I|$ defines the damping $e$-folding timescale; $\omega_I>0$, the mode is unstable and $\tau_\text{ins}\equiv 1/\omega_I$ defines the instability $e$-folding timescale; $\omega_I=0$, the mode is marginally stable or stationary. When analyzing unstable QNMs, one is commonly interested in those corresponding to the shortest instability timescales. In the present case, these are the fundamental $m=1$ QNMs.

The absence of ingoing waves at infinity is equivalent to setting $A_s^-=0$ in Eq. \eqref{eq:Wrskn3}, i.e. to requiring
\begin{equation}
    A_h^+/A_h^-=\mathcal{R}e^{-i2(\omega_\text{QNM}-mB) r_0^*}.
    \label{QNM-generic-R-Am}
\end{equation}
One can solve Eq. \eqref{QNM-generic-R-Am} for $\omega_\text{QNM}$, which yields
\begin{align}
\label{Vilken-omegaR}
\omega_R&=mB+\frac{1}{2r_0^*}\left[\arg(\mathcal{R})+\arg(A_h^-/A_h^+\right)]\\
\label{Vilken-omegaI}
\omega_I&=-\frac{1}{4r_0^*}\left(\log|\mathcal{R}|^2+\log|A_h^-/A_h^+|^2\right).
\end{align}
$\arg(\mathcal{R})$ dictates the difference in phase between ingoing and outgoing waves. If $\mathcal{R}$ is a positive (negative) real number, the phase difference is an even (odd) multiple of $\pi$. Thus, NBCs (DBCs) refer to waves reflected in phase (antiphase). If $\mathcal{R}$ is a complex number, the phase difference is a multiple of some real number between $0$ and $\pi$. Without loss of generality (as far as QNM stability is concerned), the reflectivity $\mathcal{R}$ will hereafter be considered a real parameter. Note that $\omega_R$ does not depend on the magnitude of $\mathcal{R}$. Thus, once $\arg(\mathcal{R})$ is fixed, the introduction of dissipation ($|\mathcal{R}|^2<1$) does not affect the real part of the QNM frequency. On the contrary, the imaginary part changes with changing $|\mathcal{R}|^2$. It follows from Eq. \eqref{Vilken-omegaI} that $|\mathcal{R}|^2$ determines QNM stability. Such stability is achieved whenever $\omega_I<0$, i.e. when
\begin{align}
\label{wI-stability-x}
|\mathcal{R}|^2<\left|\frac{A_h^+}{A_h^-}\right|^2=\left|\frac{A_\infty^-}{A_\infty^+}\right|^2
\end{align}
where the last equality follows from the Wronskian relations in Eqs. \eqref{eq:Wrskn1} and \eqref{eq:Wrskn2}. The last term in Eq. \eqref{wI-stability-x} is the inverse of the superradiant coefficient for the draining bathtub ($\mathcal{R}=0$). Thus, one can write
\begin{align}
\label{wI-stability}
|\mathcal{R}|^2<\frac{1}{1+Z_0(\omega_R)},
\end{align}
where $Z_0(\omega_R)\equiv Z(\omega_R,0)$. In other words, the upper bound on the range of values $|\mathcal{R}|^2$ can  take to assure stability is a function of the amplification factors for the draining bathtub only. The same result can be derived from a «bounce-and-amplify» argument \cite{BritoCardosoPani:2015,Maggio:2018}. When $|\mathcal{R}|^2=1$, the condition in Eq. \eqref{wI-stability} is satisfied only when $Z_0<0$, i.e. when the real part of the QNM frequencies does not lie in the superradiant regime ($0<\omega_R<mB$), otherwise instabilities are triggered.

\section{\label{sec:level4}Numerical results}

\subsection{Numerical method}

The numerical results to be presented in the following were obtained using a direct-integration method. The integration of Eq. \eqref{eq:RadODE-H} is performed using the numerically convenient expansions 
\begin{align*}
\tilde{H}_h(r)&=H_h(r,\varpi)+\underline{\mathcal{R}} H_h(r,-\varpi),\\
\tilde{H}_\infty(r)&=A_s^+H_\infty(r,\omega)+A_s^-H_\infty(r,-\omega)
\end{align*} 
for the radial function $H_{\omega m}(r)$ in Eq. \eqref{QNM-system-sol-in-up} in the near and in the far regions, respectively, where 
\begin{align*}
H_h(r,\varpi)&=(r-r_H)^{-i\varpi/2}\sum_{n=0}^{N_h}c_n(r-r_H)^n,\\
H_\infty(r,\omega)&=e^{+i\omega r_*}\sum_{n=0}^{N_\infty}~d_n r^{-n}.
\end{align*}
Note that $\tilde{H}_h(r_0)\approx(1+\underline{\mathcal{R}}\delta^{+i\varpi})\delta^{-i\varpi/2}$. According to Eq. \eqref{QNM-system-sol-in-up}, $H_{\omega m}(r_0)\sim(1+\mathcal{R})e^{-i\varpi r_0^*}$, meaning that one should require\footnote{Using Eq. \eqref{eq:tortoise}, one can show that $e^{-i\varpi r_0^*}=e^{-i\varpi r_0}\left(\frac{\delta}{\delta+2}\right)^{-i\varpi/2}\sim \delta^{-i\varpi/2}$, where the last step holds as long as $\delta\ll 1$.} $\tilde{H}_h(r_0)\sim(1+\mathcal{R})\delta^{-i\varpi/2}$, from which follows that $\underline{\mathcal{R}}=\mathcal{R}\delta^{-i\varpi}$. $N_h$ and $N_\infty$ are the number of terms of the partial sums. The coefficients $c_n$ and $d_n$ are functions of $\varpi$ and $\omega$, respectively, and both depend on $B$ and $m$. Inserting each expansion into Eq. \eqref{eq:RadODE-H} and equating coefficients order by order, it is possible to write $c_{1},\ldots,c_{N_h}$ ($d_{1},\ldots,d_{N_\infty}$) in terms of $c_{0}$ ($d_0$) . The latter is usually set to $1$. The choice of $N_h$ and/or $N_\infty$ should be a trade-off between computational time and accuracy. If the goal is to compute QNM frequencies, one assigns a guess value to $\omega$ and integrates Eq. \eqref{eq:RadODE-H} from $r=r_\infty$ to $r=r_0$ using the ansatz $\tilde{H}_\infty$ with $A_s^-=0$ so that the solution satisfies the relations $\tilde{H}_\infty=H_{\omega m}$ and $\D\tilde{H}_\infty/\D r=\D H_{\omega m}/\D r$ at $r=r_\infty$, where $r_\infty$ stands for the numerical value of infinity. The previous step is repeated for different guess values of $\omega$ until the solution satisfies the BC in Eq. \eqref{QNM-generic-R-Am} (one-parameter shooting). If the algorithm is numerically stable, variations in $N_\infty$ and/or $r_\infty$ do yield similar results. On the other hand, if the aim is to estimate the amplification factors defined in Eq. \eqref{eq:AmpFactors} for a given frequency $\omega$, one integrates Eq. \eqref{eq:RadODE-H} from $r=r_0$ to $r=r_\infty$ using the ansatz $\tilde{H}_h$ so that the solution satisfies the relations $\tilde{H}_h=H_{\omega m}$ and $\D\tilde{H}_h/\D r=\D H_{\omega m}/\D r$ at $r=r_0$ and then extracts the coefficients $A_s^\pm$ of the ansatz $\tilde{H}_\infty$ at infinity.

All numerical integrations were performed using the integration parameters $N_h=4$ and $N_\infty=10$. When computing QNM frequencies (amplification factors), $r_\infty$ was set to $100$ ($400$). The guess values to the QNM frequencies were chosen according to the numerical results reported in \cite{MaggioPaniFerrari:2017}.

\subsection{Quasi-normal modes}

\begin{figure*}[t]
\centering
   \subfigure{\includegraphics[scale=0.34]{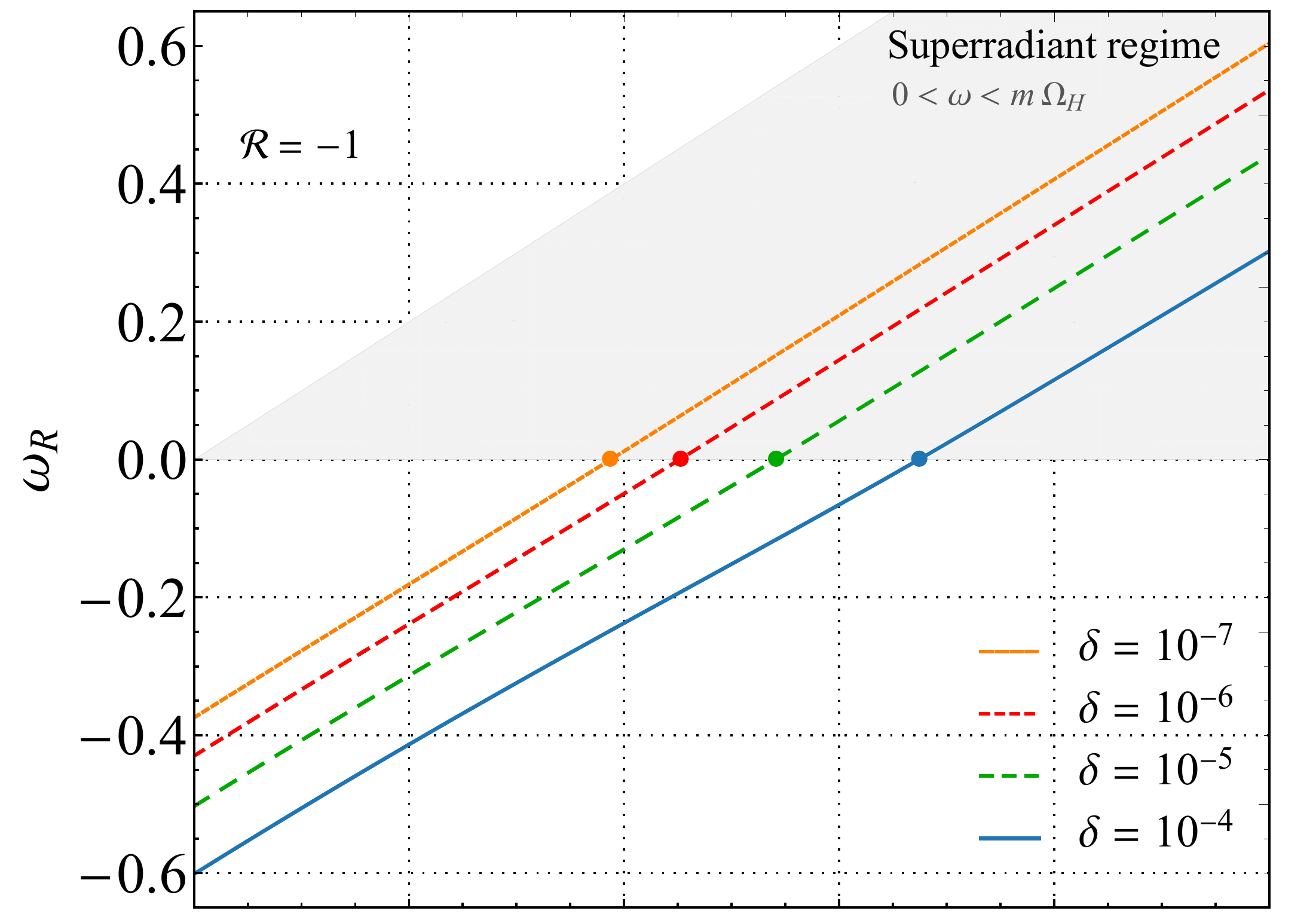}}
   \quad
   \subfigure{\includegraphics[scale=0.34]{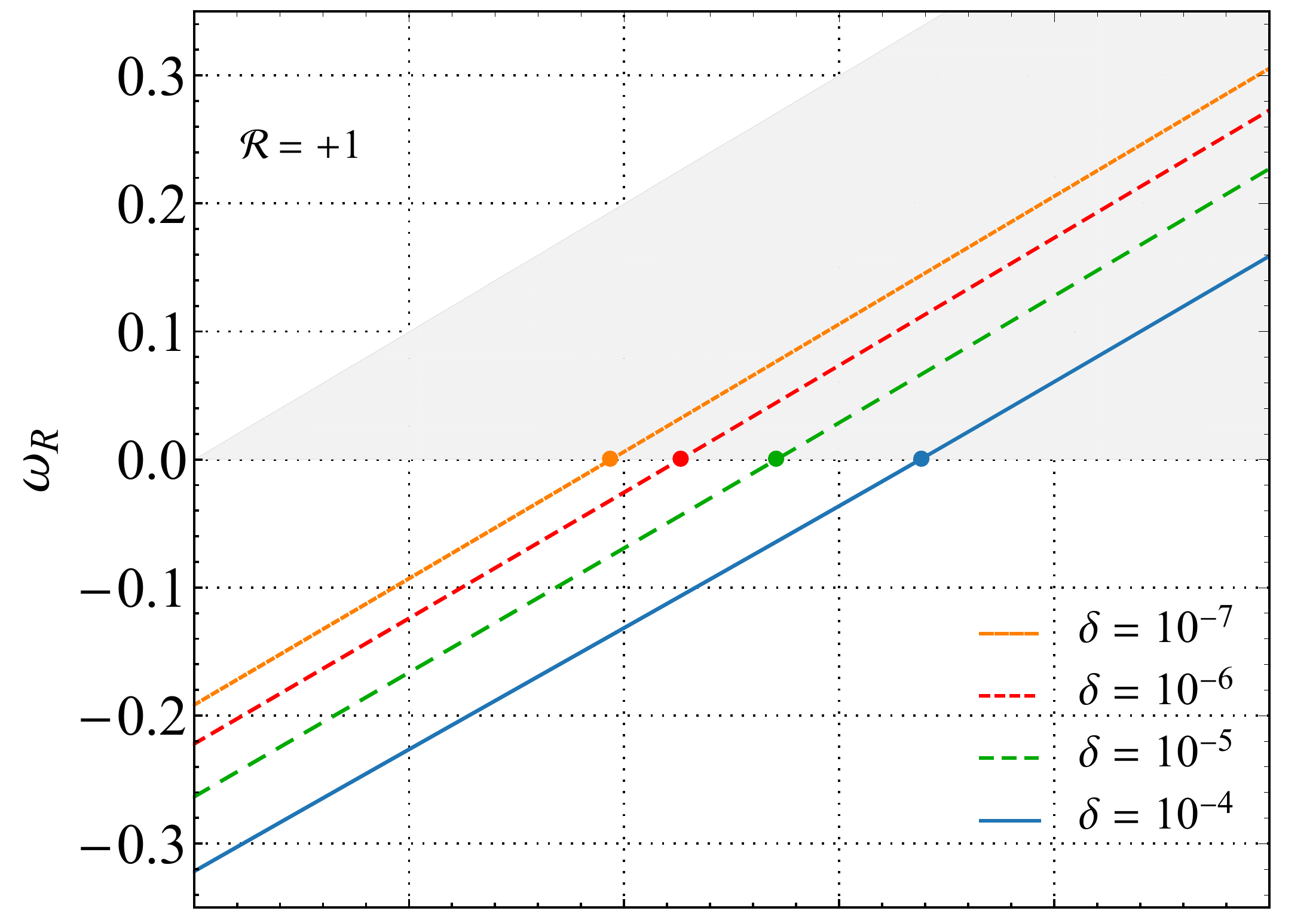}}\vspace{-0.2cm}
   \subfigure{\includegraphics[scale=0.34]{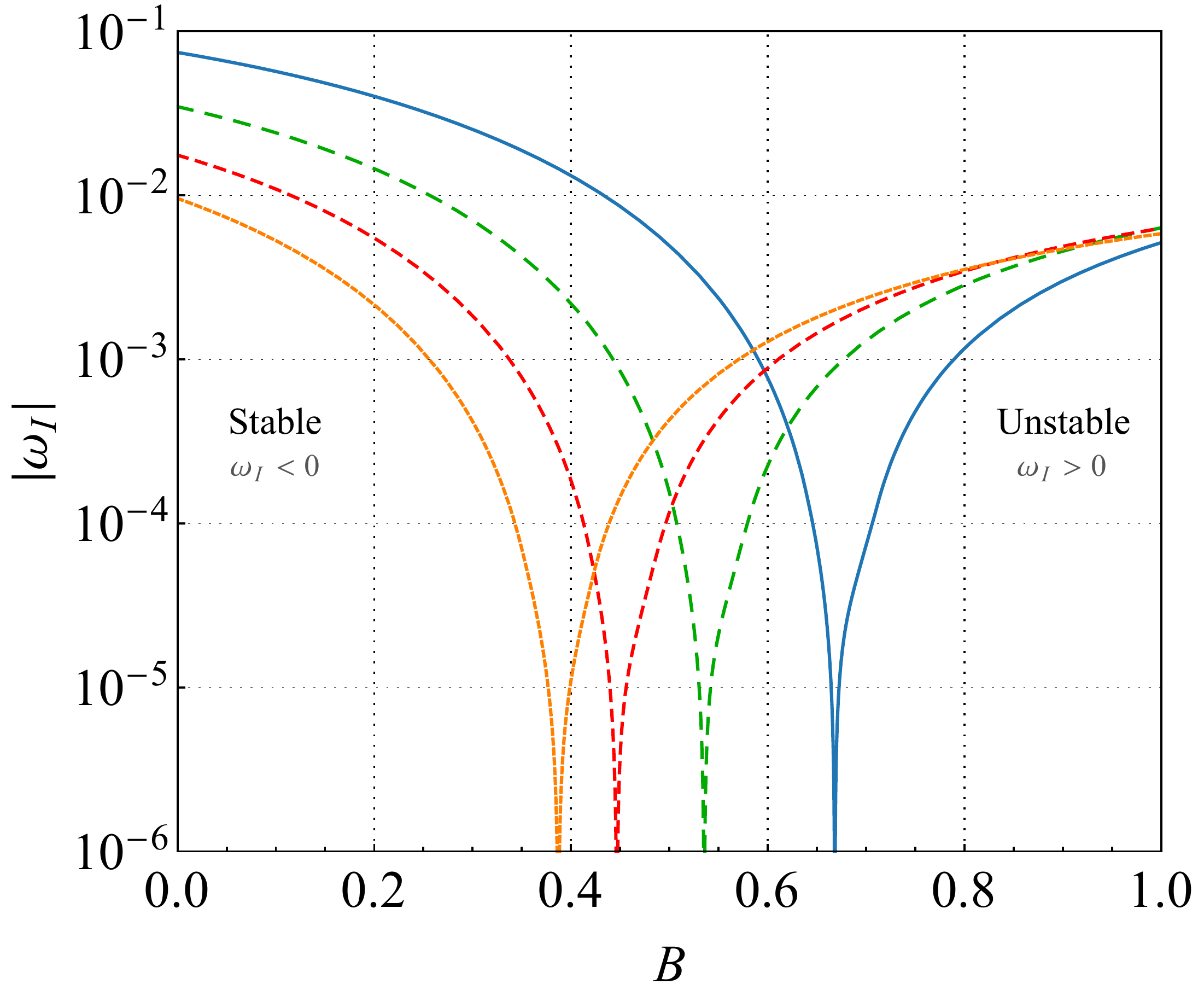}}
   \quad
   \subfigure{\includegraphics[scale=0.34]{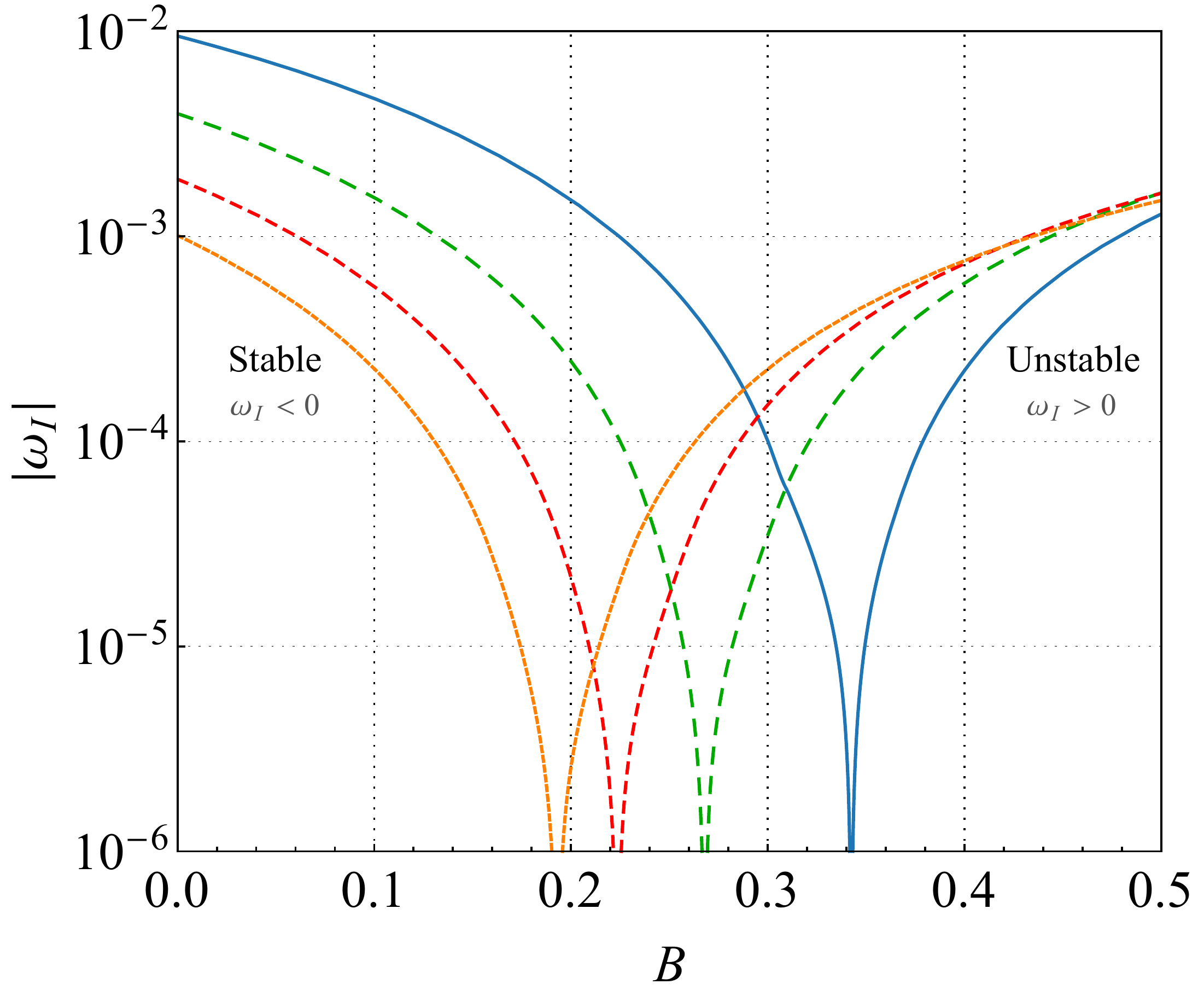}}
	\caption{Real (\textit{top}) and imaginary (\textit{bottom}) parts of the fundamental $m=1$ QNM frequencies of ECO-like vortices with a perfectly-reflecting surface at $r_0\equiv r_H(1+\delta)$, $0<\delta\ll 1$, where $r_H$ is the would-be acoustic horizon of the corresponding draining bathtub, as a function of the rotation parameter $B$, for DBCs (\textit{left}) and NBCs (\textit{right}).  The left (right) arms of the interpolating functions refer to negative (positive) frequencies. The colored dots are zero-frequency (marginally stable) QNMs.} 
\label{fig:AcousHole-QNM}
\end{figure*}

\autoref{fig:AcousHole-QNM} shows the fundamental $m=1$ QNM frequencies of perfectly-reflecting ECO-like vortices with different characteristic parameters\footnote{Since $\delta\ll 1$, $\delta$ is a more suitable parameter than $r_0$.} $\{\delta,B\}$ for DBCs ($\mathcal{R}=-1$) and NBCs ($\mathcal{R}=1$). Note that the bottom panels are plots of the absolute value of $\omega_I$. As pointed out in \autoref{sec:level2}, $\omega_R$ depends on $\arg(\mathcal{R})$ but not on $|\mathcal{R}|^2$. This means that the top panels of \autoref{fig:AcousHole-QNM} give information about the real part of the QNM frequencies of both perfectly- and partially-reflecting ECO-like vortices. 

The results are qualitatively similar for both BCs. At first order in $B$ (i.e. for $B\lesssim 0.1$), $\omega_R$ is a linear function of the angular velocity $B$, as one would expect from Eq. \eqref{Vilken-omegaR}. The initial value (corresponding to $B=0$) depends on $r_0^*$ or, more precisely, on the the inverse of $\log\delta$, in accordance with Eq. \eqref{eq:tortoise}. Similarly, $\omega_I$ grows monotonically with increasing rotation. Both $\omega_R$ and $\omega_I$ change sign from negative to positive as $B$ increases. Within numerical accuracy, the sign changes occur at the same critical value $B_c$, meaning $\omega_R,\omega_I<0$ when $B<B_c$ and $\omega_R,\omega_I>0$ when $B>B_c$. In other words, QNMs turn from stable to unstable as the fluid spins faster and faster and, furthermore, perfectly-reflecting ECO-like vortices admit zero-frequency ($\omega=0$) QNMs. Similar phenomenological aspects regarding the interaction of massless bosonic fields with perfectly-reflecting Kerr-like ECOs have been reported in \cite{MaggioPaniFerrari:2017,Hod:2017,Maggio:2018}. It was shown in particular that some ECOs or ECO analogues can only support static configurations of a scalar field for a discrete set of critical radii \cite{Hod:2017,Hod:UltraSECOs:2017,Hod:Vortex:2017,Hod:ChargedECOs:2017}. This also holds true for the present case, as shown in the Appendix \ref{apx:2}.

The aforementioned instability finds its origin in the possible existence of negative-energy physical states inside the ergoregion. In general, in BH physics, the absence of an event horizon turns horizonless rotating ECOs unstable \cite{Friedman:1978,MaggioPaniFerrari:2017}. The event horizon of Kerr BHs, which can be regarded as a perfectly-absorbing surface ($\mathcal{R}=0$), prevents the falling into lower and lower negative-energy states. The same does not occur when considering perfectly-reflecting BCs, hence the development of instabilities. 

\begin{figure}[h!]
\begin{minipage}[b]{.45\textwidth}
\centering
\includegraphics[scale=0.35]{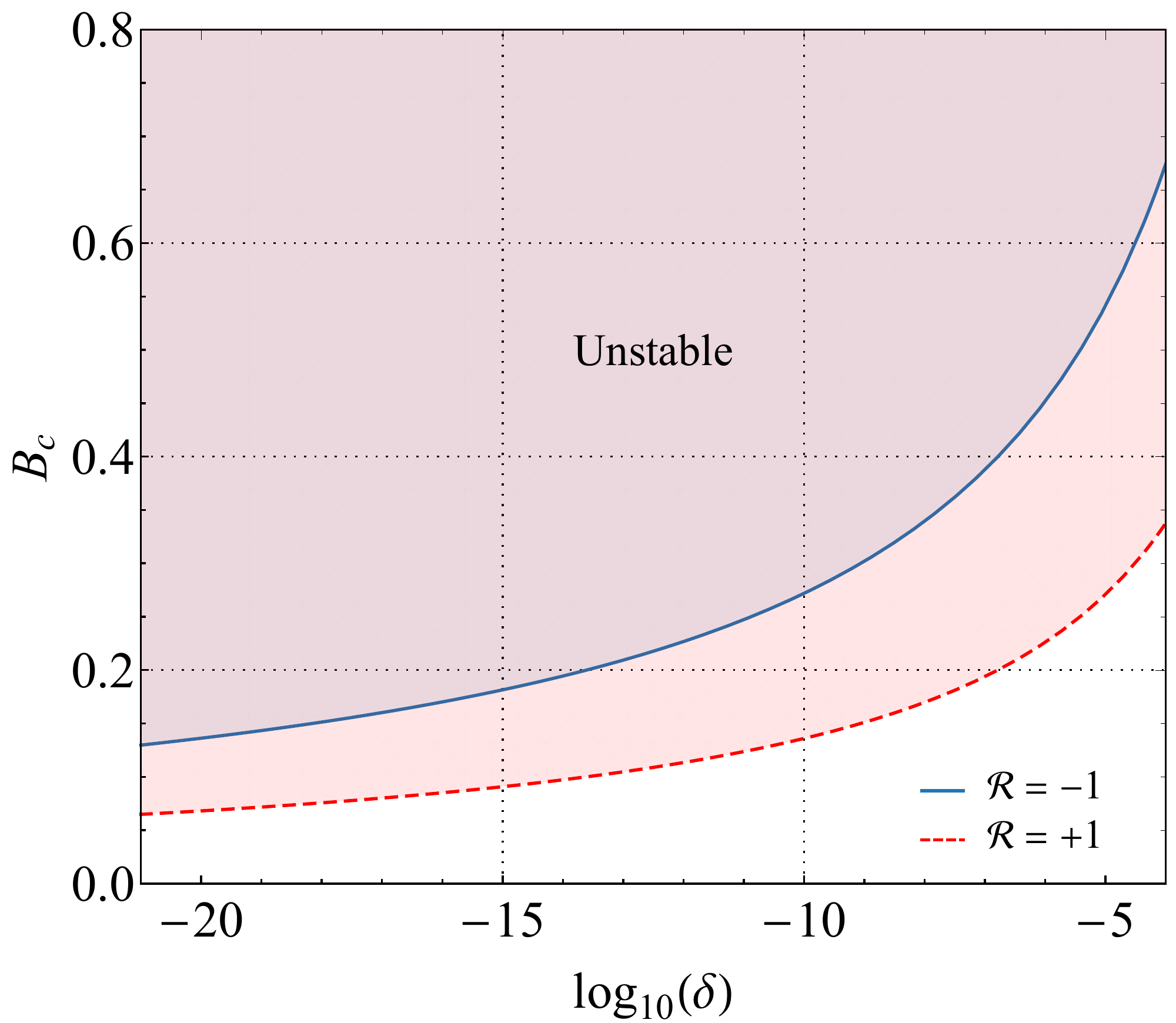}
\caption{Ergoregion instability rotation parameter threshold of the fundamental $m=1$ QNMs of perfectly-reflecting ECO-like vortices for both DBCs ($\mathcal{R}=-1$) and NBCs ($\mathcal{R}=1$). The shaded regions refer to the domain of the instability.}
\label{fig:AH-QNM-InsDom}
\end{minipage}
\end{figure}

All the positive frequencies in the top panels of \autoref{fig:AcousHole-QNM} refer to exponentially growing modes which meet the superradiance condition, that is to say that acoustic perturbations with such frequencies are amplified ($Z>0$) when scattered off perfectly-reflecting ECO-like vortices, thanks to the transfer of angular momentum and energy from the latter to the former. According to Eq. \eqref{wI-stability}, only the absence of superradiant amplification when $\mathcal{R}=0$ ($Z_0<0$) guarantees exponentially decaying responses of perfectly-reflecting ($|\mathcal{R}|^2=1$) ECOs to external linear perturbations. From a dynamical point of view, the instability is expected to result in the scatterer spinning slower and slower until its fundamental QNM mode, whose frequency depends on $B$, starts decaying rather than growing over time, which occurs when $B=B_c$ (note that $Z<0$ when $B<B_c$). The instability domain of ECO-like vortices is depiected in \autoref{fig:AH-QNM-InsDom}. The threshold decreases monotonically as $r_0\rightarrow r_H$ (i.e. as $\delta\rightarrow0$).

\begin{figure}[h]
\begin{minipage}[b]{.45\textwidth}
\centering
\includegraphics[scale=0.35]{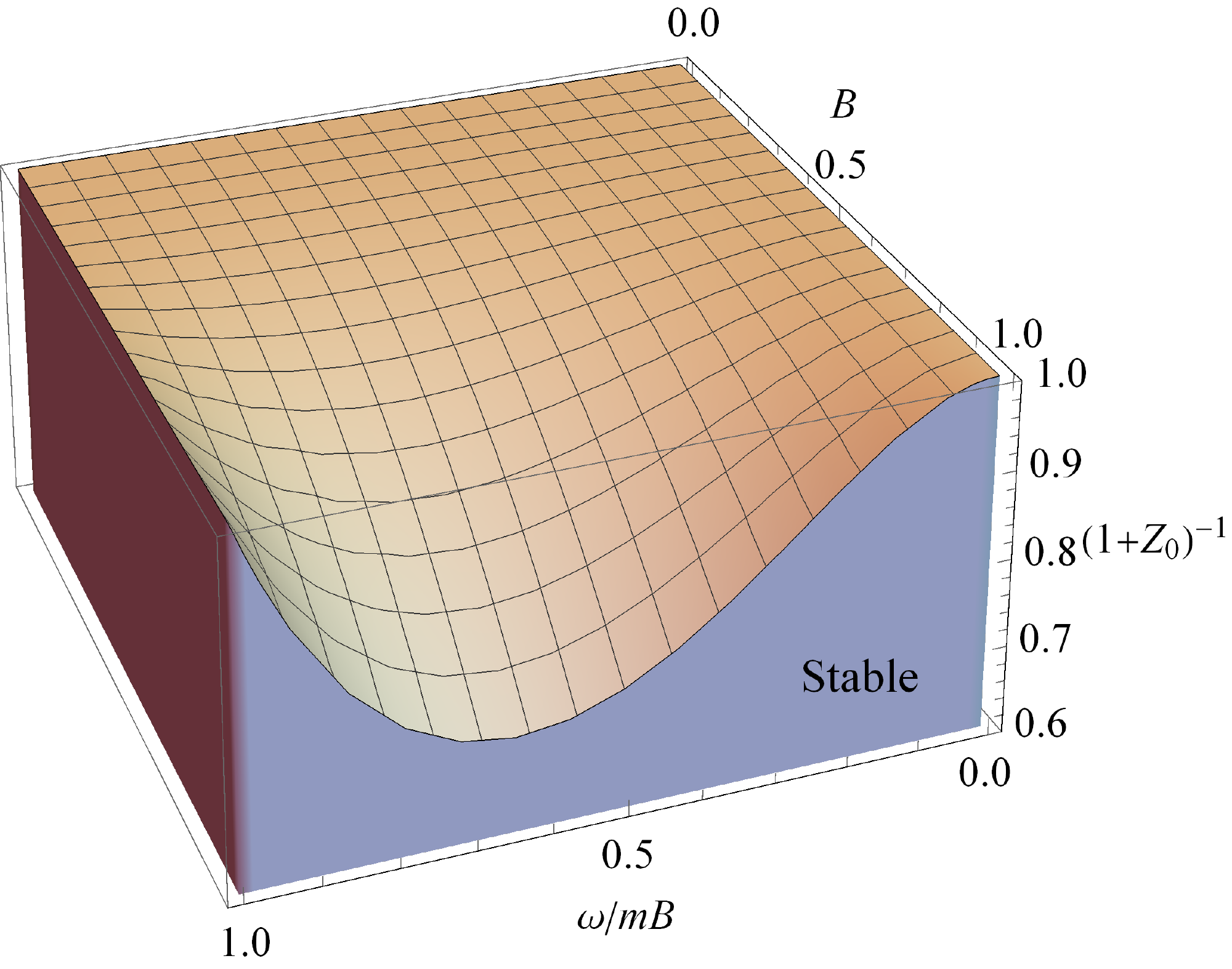}
\caption{Plot of $(1+Z_0)^{-1}$ as a function of the rotation parameter $B$ and of the frequency $\omega$ of acoustic perturbations. When $|\mathcal{R}|^2<(1+Z_0)^{-1}$, ECO-like vortices are dynamically stable. Note that the upper limit on $|\mathcal{R}|^2$ decreases as $B$ increases.}
\label{fig:AH-QNM-InsDom-B-omega}
\end{minipage}
\end{figure}

Is there any way of preventing ECO-like vortices from developing instabilities without changing $B$? The answer is affirmative. In the presence of superradiance ($Z_0>0$), unless $B<B_c$, it is clear that only partially-reflecting ECO-like vortices may be stable. However, attention must be paid to the fact that an ECO with $|\mathcal{R}|^2<1$ is not perforce dynamically stable. In fact, for each frequency satisfying the superradiance condition, there is a range of values $|\mathcal{R}|^2$ can take to assure stability. Using the small-frequency approximations in Eqs. \eqref{eq:AN-match-cond-A} and \eqref{eq:AN-match-cond-B}, Eq. \eqref{wI-stability} takes the explicit form
\begin{align}
|\mathcal{R}|^2<\left|\frac{\Gamma(m)^2+i\pi\psi_i\chi(\omega/2)^{2m}}{\Gamma(m)^2-i\pi\psi_i\chi(\omega/2)^{2m}}\right|^2,
\end{align}
where $\chi=\beta(\beta-m)^{-1}\prod_{n=0}^{m-1}(\beta-m+n)^2$. The upper bound on $|\mathcal{R}|^2$ represents the threshold for the manifestation of ergoregion instabilities and depends both on $\omega$ and $B$, as shown in \autoref{fig:AH-QNM-InsDom-B-omega}. 

One is interested in setting restrictions on $|\mathcal{R}|^2$ which are frequency-independent, in order to avert exponentially growing linear perturbations regardless of their frequency. The absolute upper bound on $|\mathcal{R}|^2$ from Eq. \eqref{wI-stability}, herein dubbed maximum reflectivity, is set by the maximum amplification factor when $\mathcal{R}=0$, i.e. to the minimum of the function $(1+Z_0)^{-1}$, and is therefore independent of $\delta$. For sufficiently slow drains ($B\leq 1$), ECO-like vortices are stable against acoustic perturbations of any frequency as long as $|\mathcal{R}|^2\lesssim 70\%$.

\subsection{Superradiant scattering}

\begin{figure*}[th]
\centering
   \subfigure{\includegraphics[scale=0.33]{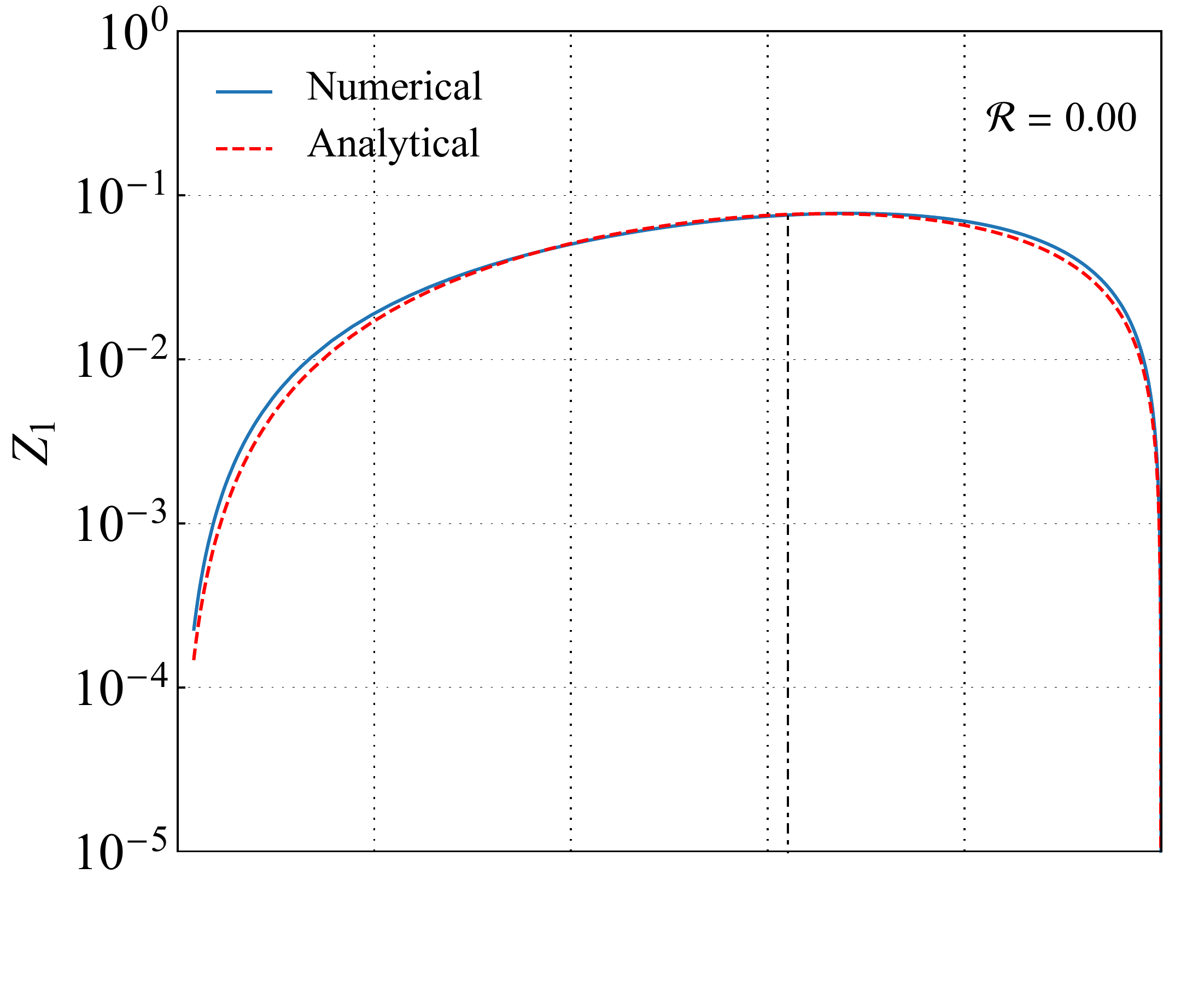}}\hspace{-0.9cm}
   \subfigure{\includegraphics[scale=0.33]{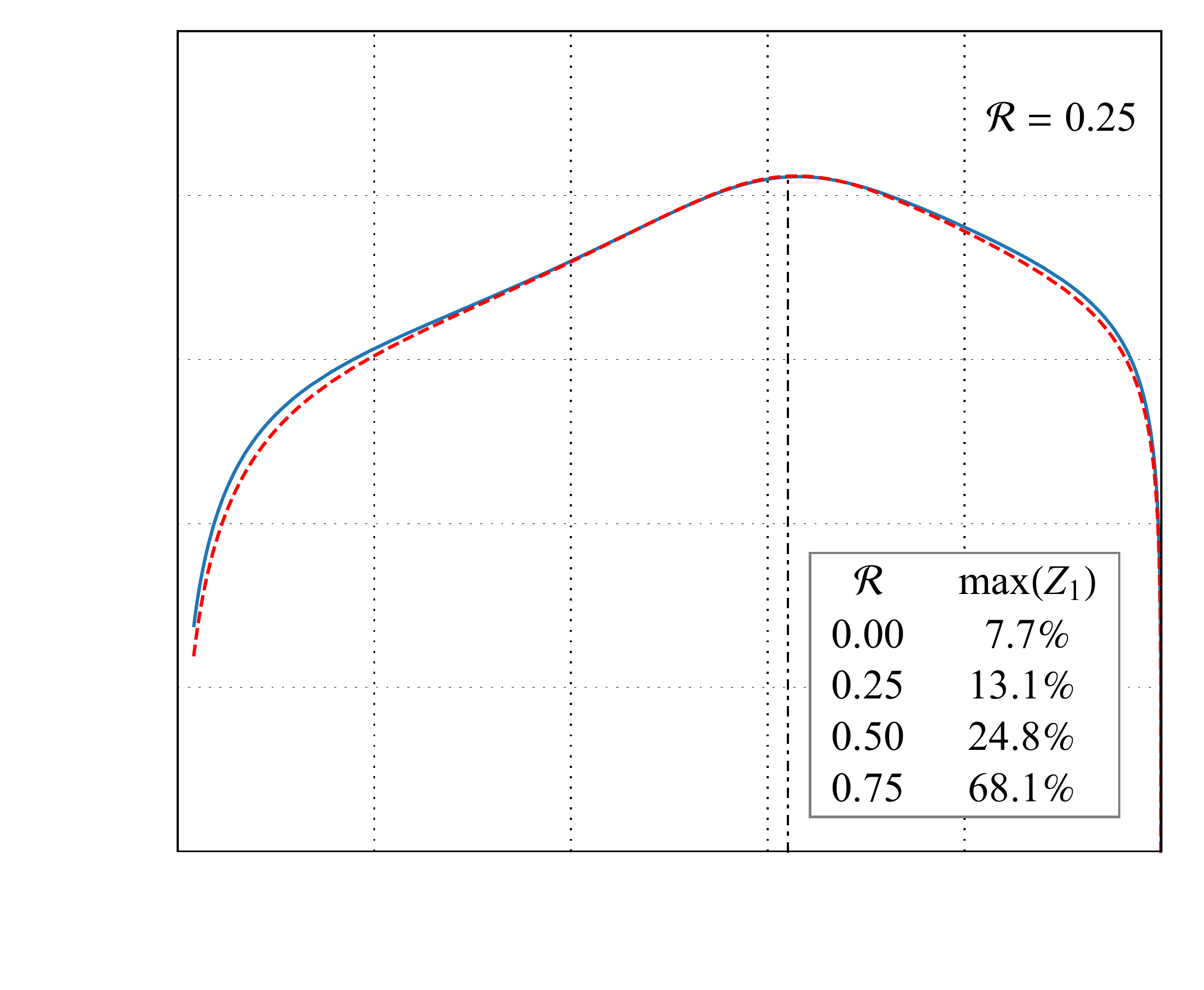}}\vspace{-0.9cm}
   \subfigure{\includegraphics[scale=0.33]{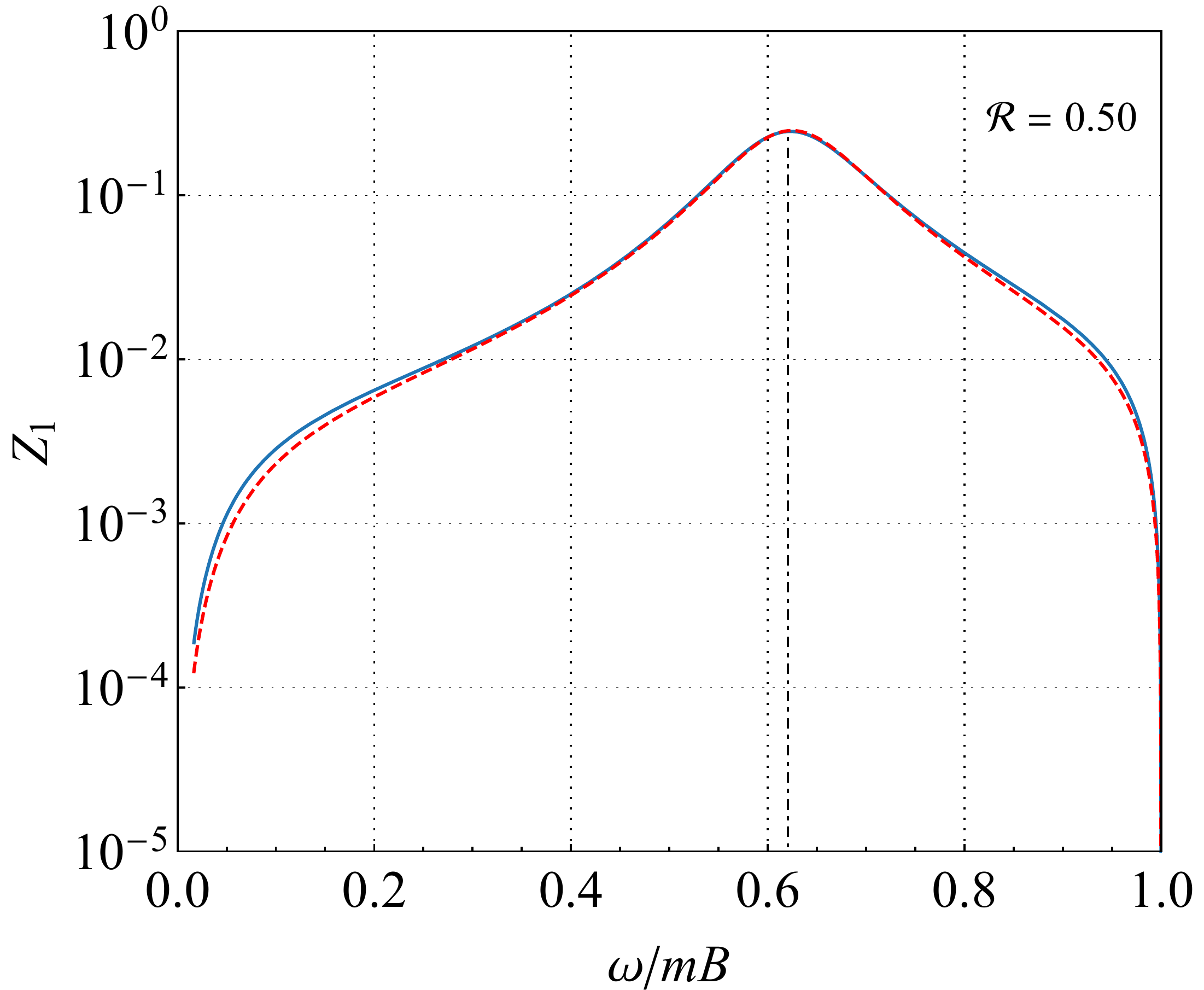}}\hspace{-0.9cm}
   \subfigure{\includegraphics[scale=0.33]{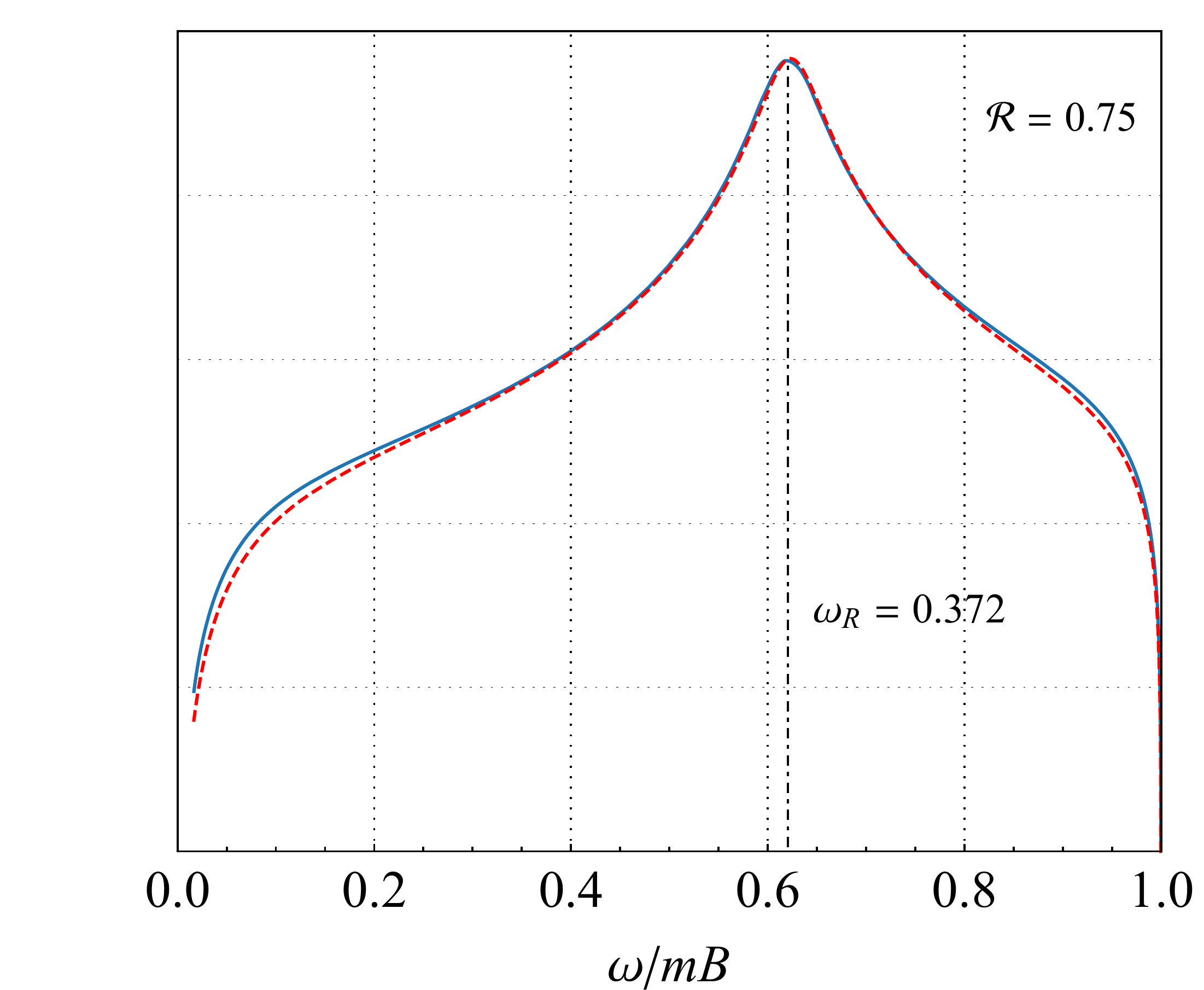}}
	\caption{Numerical and analytical values of the amplification factors for superradiant ($0<\omega<mB$) acoustic perturbations with $m=1$ scattered off a ECO-like vortex with $B=0.6$ and featuring a surface with reflectivity $\mathcal{R}$ at $r_0= r_H(1+\delta)$, where $r_H$ is the would-be acoustic horizon of the corresponding draining bathtub and $\delta=10^{-6}$. The resonance matches the real part of the fundamental QNM frequency of the vortex-like ECO (dotted vertical lines).} 
	\label{fig:ECO-Vortices-AmpFac}
\end{figure*}

\begin{figure}[h!]
\centering
   \subfigure{\includegraphics[scale=0.33]{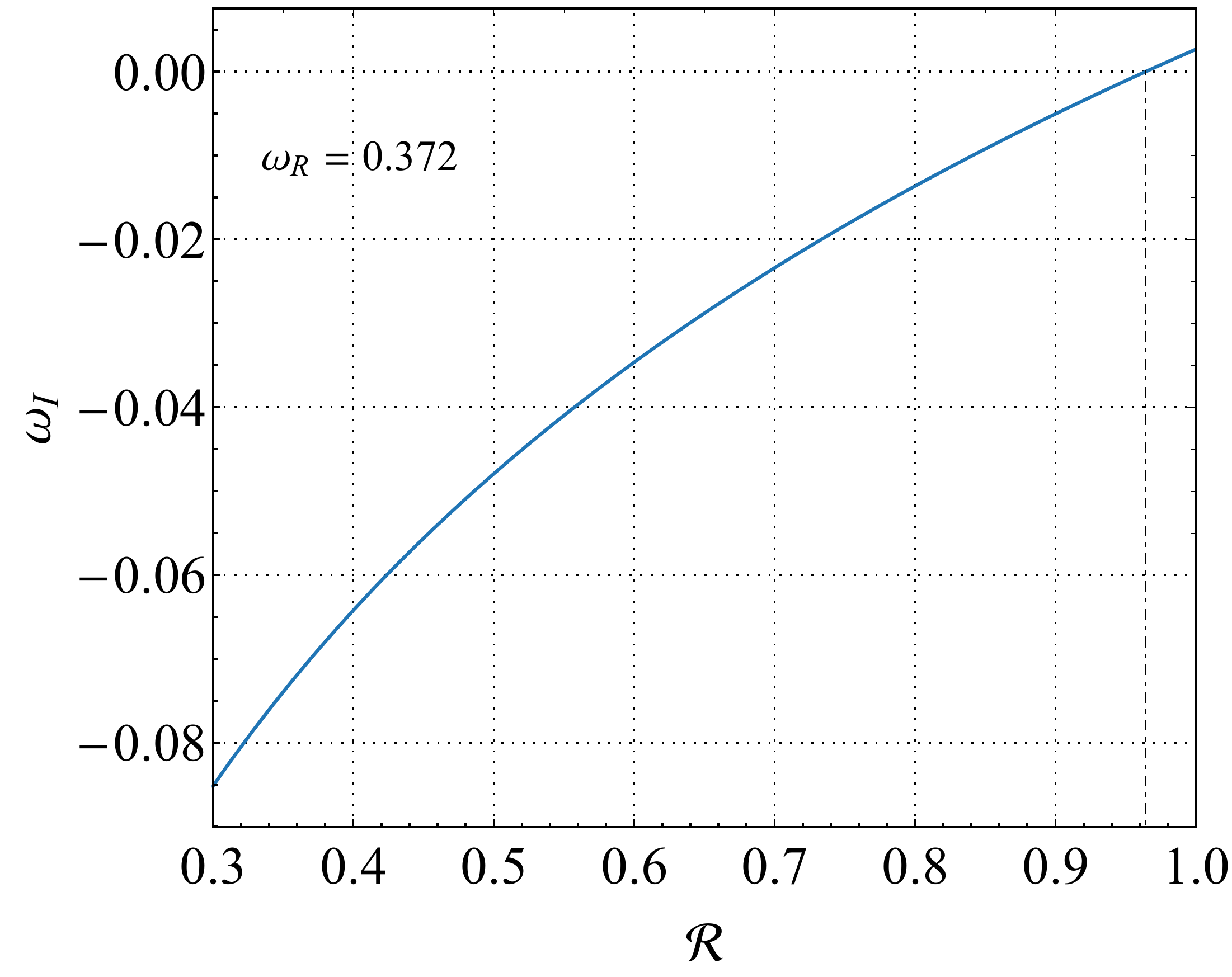}}
	\caption{Imaginary part of the QNM frequency which sets the peaks of the amplification factors plotted in \autoref{fig:ECO-Vortices-AmpFac}, as a function of the reflectivity $\mathcal{R}$. $\omega_I$ vanishes at $\mathcal{R}\approx 0.964$ (dotted vertical line).\\} 
	\label{fig:fig1}
\end{figure}

\begin{figure}[h!]
\vspace{-0.2cm}
\centering
   \subfigure{\includegraphics[scale=0.33]{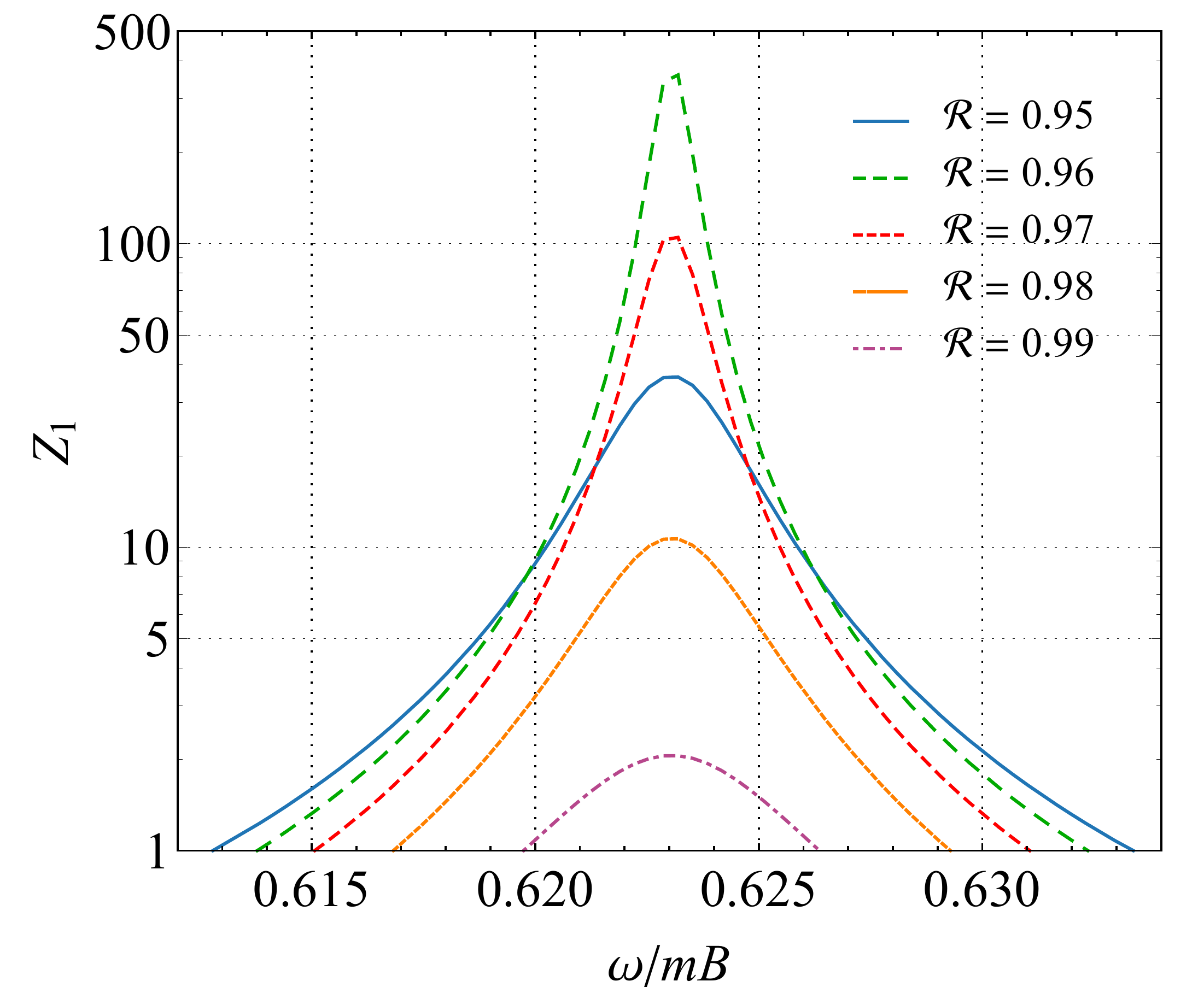}}
	\caption{Analytical approximation for the amplification factors plotted in \autoref{fig:ECO-Vortices-AmpFac} in the neighborhood of the fundamental $m=1$ QNM frequency for reflectivities near $\mathcal{R}\approx0.964$.} 
	\label{fig:fig2}
\end{figure}

\autoref{fig:ECO-Vortices-AmpFac} illustrates the amplification factors for superradiant $m=1$ acoustic linear perturbations scattered off ECO-like vortices with $\delta=10^{-6}$, $B=0.6$ and different reflectivities. The numerical results were obtained via the direct-integration method described previously, whereas the analytical ones were computed using the small-frequency approximations in Eqs. \eqref{eq:AN-match-cond-A} and \eqref{eq:AN-match-cond-B} via Eq. \eqref{eq:AmpF}. The closeness between numerical and analytical results is evident. The data for different reflectivities share some qualitatively features: the amplification factors appear to vanish at the endpoints of the superradiant regime (i.e. at $\omega=0$ and $\omega=mB$) and have a maximum value. The graph referring to the draining bathtub ($\mathcal{R}=0$) is in agreement with numerical results previously reported in the literature \cite{CardosoLemosYoshida:2004,CrispinoDolanOliveira:2012}. The maximum amplification is about $8\%$ when $\mathcal{R}=0$, meaning the maximum reflectivity is approximately $92\%$ when $B=0.6$.  When $|\mathcal{R}|^2$ is nonvanishing, a resonance becomes noticeable around a frequency of about $0.372$ (dotted vertical lines in \autoref{fig:ECO-Vortices-AmpFac}), which matches the real part of a QNM frequency precisely. Like in classical mechanics, an acoustic linear perturbation extracts more angular momentum and energy when its frequency coincides with the object’s proper frequencies of vibration. The peak's height is determined by the imaginary part of $\omega_\text{QNM}$ \cite{Macedo:2018}, being maximum when $\omega_I=0$. As shown in \autoref{fig:fig1}, the QNM which sets the peaks in \autoref{fig:ECO-Vortices-AmpFac} is marginally-stable when $\mathcal{R}\approx0.964$ (dotted vertical line in \autoref{fig:fig1}). \autoref{fig:fig2} confirms that the maximum value of the amplification factor occurs indeed for a reflectivity around $0.964$.

\section{\label{sec:level6}Conclusion}

GW astronomy opens a new window on the universe and is expected to unveil spacetime features in the vicinity of compact objects, testing both general relativity and BH physics predictions. However, present GW observations are not precise enough to (indirectly) probe the true nature of BH candidates and do not rule out alternative scenarios. This has been one of the strongest motivations behind ECO models. Their phenomenology has been widely addressed in search of alternatives to the BH paradigm. 

Following this trend, this work aimed to explore the phenomenology of acoustic perturbations of an analogue model for ECOs built from the draining bathtub geometry, named herein ECO-like vortices. These objects feature an ergoregion and are endowed with a surface with reflective properties rather than an acoustic horizon. Although ECO-like vortices do have the key ingredients to trigger ergoregion instabilities, it turns out that dissipation mitigates or even neutralizes exponentially growing modes.

The analysis led to two main conclusions, supported by a low-frequency analysis and by direct-integration numerical calculations. First, when the object's surface is perfectly-reflecting $(|\mathcal{R}|^2=1)$, an instability develops when the vortex is spinning at a rate above some critical value of the rotation parameter. Despite the dependence of the instability domain on the location of the surface, it generally occurs when $B>\mathcal{O}(0.1)$. The instability is intimately linked to the ergoregion, where negative-energy physical states can form. These cannot be absorbed by the vortex's surface and, therefore, cause the exponential growth of acoustic perturbations. The ergoregion instability of ECO-like vortices is attenuated or neutralized when its surface is not perfectly- but partially-reflecting. An absorption coefficient greater than approximately $30\%$ prevents unstable QNMs to develop in ECO-like vortices with $B$ below unit. The results are similar to those reported in \cite{MaggioPaniFerrari:2017,Maggio:2018,Macedo:2018} and attests that a general way of preventing such instabilities from arising is to incorporate dissipative-like effects into the surface. 

Second, the stimulation of exponentially growing QNMs is optimized for more reflective surfaces and, therefore, is expected to generate narrower spectral lines in the emission cross sections, similarly to those in the absorption cross sections of spherically symmetric ECOs \cite{Macedo:2018}. An interesting extension of the work presented herein would precisely be to compute the emission cross sections of ECO-like vortices, as it would be useful in probing the effects of dissipation upon  superradiant scattering in possible future experiments. Implementing a setup which reproduces the ECO-like vortex introduced here appears to be a thorny issue. This would require to place a right circular cylinder at the acoustic horizon of a vortex flow. 

Moreover, the reflectivity $\mathcal{R}$ was assumed to be frequency-independent. A possible future extension may lift this assumption and consider ECOs with frequency-dependent reflectivities. 

\vfill

\begin{acknowledgments}
This work has been supported by Funda\c{c}\~ao para a Ci\^encia e a Tecnologia (FCT) grant PTDC/FIS-OUT/28407/2017 and by CENTRA (FCT) strategic project UID/FIS/00099/2013.   This work has further been supported by  the  European  Union's  Horizon  2020  research  and  
innovation  (RISE) programmes H2020-MSCA-RISE-2015
Grant No.~StronGrHEP-690904 and H2020-MSCA-RISE-2017 Grant No.~FunFiCO-777740. The authors would like to acknowledge networking support by the COST Action CA16104.
\end{acknowledgments}

\appendix
\section{Analytical analysis}
\label{sec:level3}

A number of relevant quantities, such as QNM frequencies and amplification factors, can be computed analytically in the low-frequency regime ($\omega\ll 1$) using matching-asymptotic techniques \cite{CardosoPaniCadoniCavaglia:2008}. Contrarily to previous works \cite{Cardoso:2006}, mostly focused on the BH case ($\mathcal{R}=0$), here the presence of outgoing waves near the would-be acoustic horizon is considered and the problem is solved for a generic reflectivity $\mathcal{R}$. For that purpose, the spacetime region outside the reflective surface at $r=r_0$ is split into a region \textit{near} the would-be acoustic horizon, i.e. in the limit $r-r_H\ll 1/\omega$, and a region \textit{far} from it, i.e. at infinity, where $r\gg r_H$. In the following, besides the condition $B\omega\ll 1$, the assumption $\varpi\ll 1$ for frequencies in the superradiant regime is also considered. One starts looking for asymptotic solutions to Eq. \eqref{eq:RadODE-S} in each spacetime region and imposing the BC in Eq. \eqref{QNM-generic-R-Am}, and then matches them in the overlapping region, where $1\ll r-r_H\ll 1/\omega$. 

To solve Eq. \eqref{eq:RadODE-S} in the far region, it is convenient to rewrite it in the form 
\begin{equation}
\label{eq:RadODE-S-2}
\left[\Delta\frac{\D}{\D r}\left(\Delta\frac{\D}{\D r}\right)+\left(\omega r-\frac{mB}{r}\right)^2-\frac{\Delta m^2}{r}\right]S_{\omega m}(r)=0,
\end{equation}
where $\Delta(r)=rg(r)$. When $r\gg r_H$ and $B\omega\ll 1$, Eq. \eqref{eq:RadODE-S-2} reduces to a Bessel ODE,
\begin{equation}
\left[r^2\frac{\D^2}{\D r^2}+r\frac{\D}{\D r}+(\omega^2r^2-m^2)\right]S_{\omega m}(r)=0,
\end{equation}
whose general solution is a linear combination of Bessel functions of the first and second kinds,
\begin{equation}
\label{eq:RadR-INF-large}
S_{\omega m}(r)=\mathscr{A}J_m(\omega r)+\mathscr{B}Y_m(\omega r),
\end{equation}
where $\mathscr{A},\mathscr{B}\in\mathbb{C}$. The large-$r$ behavior of the asymptotic solution in Eq. \eqref{eq:RadR-INF-large} is
\begin{equation}
S_{\omega m}(r)\sim\sqrt{\frac{2}{\pi\omega r}}\left[\mathscr{A}\cos(\omega r-\varsigma)+\mathscr{B}\sin(\omega r-\varsigma)\right],
\end{equation}
where $\varsigma\equiv\pi(m+1/2)/2$, which can be written in terms of the radial function $H_{\omega m}(r)$ and as a superposition of ingoing and outgoing waves. In effect,
\begin{equation}
\label{eq:Rad-INF-large-e}
H_{\omega m}(r)\sim\frac{(\mathscr{A}+i\mathscr{B})e^{-i(\omega r-\varsigma)}+(\mathscr{A}-i\mathscr{B})e^{+i(\omega r-\varsigma)}}{\sqrt{2\pi\omega}}.
\end{equation}
The amplitudes of the ingoing and outgoing waves are proportional to $(\mathscr{A}+i\mathscr{B})$ and $(\mathscr{A}-i\mathscr{B})$, respectively. The main goal of the asymptotic matching is to find expressions for $\mathscr{A}$ and $\mathscr{B}$ in terms of $r_0$ (or $\delta$), $\mathcal{R}$, $B$, $m$ and $\omega$.

The small-$r$ behavior of the asymptotic solution in Eq. \eqref{eq:RadR-INF-large} is
\begin{equation}
\label{eq:RadS-INF-small}
S_{\omega m}(r)\sim\mathscr{A}\frac{(\omega/2)^m}{\Gamma(m+1)}r^{+m}-\mathscr{B}\frac{\Gamma(m)}{\pi(\omega/2)^{m}}r^{-m}.
\end{equation}

In the near region, by defining
\begin{equation}
\label{eq:RadFun-ST}
S_{\omega m}(r)=r^{-m}\left[g(r)\right]^{-i\frac{\varpi}{2}}T_{\omega m}(r),
\end{equation}
introducing the new coordinate $z=r^{-2}$ and using the conditions $B\omega\ll 1$ and $\varpi\ll 1$, one can bring Eq. \eqref{eq:RadODE-S} into the hypergeometric form
\begin{equation}
\label{eq:RadODE-T}
\left[z(1-z)\frac{\D^2}{\D z^2}+[\gamma-(\alpha+\beta+1)z]\frac{\D}{\D z}-\alpha\beta\right]T_{\omega m}(z)=0,
\end{equation}
where $2\alpha=2+m-i\varpi$, $2\beta=m-i\varpi$ and $\gamma=m+1$. It is easy to check that $\alpha$, $\beta$ and $\gamma$ satisfy the relations $\alpha-\beta=1$, $\alpha+\beta=2\alpha-1=2\beta+1$ and $\gamma-\alpha-\beta=i\varpi$.

The hypergeometric differential equation has three singular points: $z=0,1,\infty$. Its most general solution in the neighborhood of $z=1$ reads
\begin{align}
\label{eq:RadH-HOR-small}
T_{\omega m}(z)=\mathscr{C}T_{\omega m}^\text{i}(z)+\mathscr{D}(1-z)^{\gamma-\alpha-\beta}T_{\omega m}^\text{o}(z),
\end{align}
where $\mathscr{C},\mathscr{D}\in\mathbb{C}$, $T_{\omega m}^\text{i}(z)={}_2F_1(\alpha,\beta;\alpha+\beta-\gamma+1;1-z)$ and $T_{\omega m}^\text{o}(z)={}_2F_1(\gamma-\beta,\gamma-\alpha;\gamma-\alpha-\beta+1;1-z)$. ${}_2F_1(\alpha,\beta;\gamma;z)$ is the hypergeometric function.
Since $T_{\omega m}^\text{i}(1)=T_{\omega m}^\text{o}(1)=1$, one finds that the small-$r$ behavior of the asymptotic solution in Eq. \eqref{eq:RadH-HOR-small} is
\begin{align}
T_{\omega m}(z)\sim\mathscr{C}+\mathscr{D}(1-z)^{+i\varpi},
\end{align}
Using Eqs. \eqref{eq:RadFun-SH} and \eqref{eq:RadFun-ST}, one can show that
\begin{align}
H_{\omega m}(r)&\sim\underline{\mathscr{C}}e^{-i\varpi r_*}+\underline{\mathscr{D}}e^{+i\varpi r_*}
\end{align}
near the would-be acoustic horizon, where
\begin{align}
\underline{\mathscr{C}}\equiv\mathscr{C}\left(\frac{r+1}{re^r}\right)^{-i\varpi},
\quad
\underline{\mathscr{D}}\equiv\mathscr{D}\left(\frac{r+1}{re^r}\right)^{+i\varpi}.
\end{align}
Therefore, the BC \eqref{QNM-generic-R-Am} turns into
\begin{equation}
\label{eq:ABR}
\frac{\mathscr{D}}{\mathscr{C}}=\mathcal{R}\left(\frac{r_0^2}{r_0^2-1}\right)^{i\varpi}.
\end{equation}

Given that $\gamma\in\mathbb{Z}$, one must be careful when analyzing the large-$r$ (small-$z$) behavior of the asymptotic solution \eqref{eq:RadH-HOR-small} \cite{HandbookMF,SFMathPhys}. One can show that \cite{Cardoso:2006}
\begin{align*}
T_{\omega m}^\text{i}(z)&\sim\mathscr{T}_\text{i}\left[\frac{(-1)^{m-1}}{m(\alpha-m)_m(\beta-m)_m}z^{-m}+\psi_\text{i}\right],\\
T_{\omega m}^\text{o}(z)&\sim\mathscr{T}_\text{o}\left[\frac{(-1)^{m-1}}{m(1-\alpha)_m(1-\beta)_m}z^{-m}+\psi_\text{o}\right],
\end{align*}
where
\begin{align*}
\mathscr{T}_\text{i}&\equiv\frac{(-1)^{m+1}}{m!}\frac{\Gamma(\alpha+\beta-m)}{\Gamma(\alpha-m)\Gamma(\beta-m)},\\
\psi_\text{i}&\equiv\psi(\alpha)+\psi(\beta)-\psi(m+1)-\psi(1),\\
\mathscr{T}_\text{o}&\equiv\frac{(-1)^{m+1}}{m!}\frac{\Gamma(m+2-\alpha-\beta)}{\Gamma(1-\alpha)\Gamma(1-\beta)},\\
\psi_\text{o}&\equiv\psi(m+1-\alpha)+\psi(m+1-\beta)-\psi(m+1)-\psi(1).
\end{align*}
$(q)_n$ is the rising Pochhammer symbol and $\psi(\cdot)$ is the digamma function \cite{HandbookMF,SFMathPhys}. 

Using Eq. \eqref{eq:RadFun-ST} and rearranging the terms, one gets
\begin{equation}
\label{eq:RadH-HOR-large}
S_{\omega m}(r)\sim\underline{\mathscr{A}}r^{+m}+\underline{\mathscr{B}}r^{-m},
\end{equation}
where
\begin{align*}
&\underline{\mathscr{A}}\equiv\frac{(-1)^{m-1}}{m}\left[\frac{\mathscr{T}_\text{i}\mathscr{C}}{(\alpha-m)_m(\beta-m)_m}+\frac{\mathscr{T}_\text{o}\mathscr{D}}{(1-\alpha)_m(1-\beta)_m}\right],\\
&\underline{\mathscr{B}}\equiv\psi_\text{i}\mathscr{T}_\text{i}\mathscr{C}+\psi_\text{o}\mathscr{T}_\text{o}\mathscr{D}.
\end{align*}
Eq. \eqref{eq:RadH-HOR-large} exhibits the same dependence on $r$ as Eq. \eqref{eq:RadS-INF-small}. Matching the two solutions, it is straightforward to show that
\begin{align}
\label{eq:AN-match-cond-A}
\mathscr{A}&=\frac{\Gamma(m+1)}{(\omega/2)^m}\underline{\mathscr{A}},\\
\label{eq:AN-match-cond-B}
\mathscr{B}&=-\frac{\pi(\omega/2)^m}{\Gamma(m)}\underline{\mathscr{B}}.
\end{align}
One can derive some relevant physical quantities from Eqs. \eqref{eq:AN-match-cond-A}, \eqref{eq:AN-match-cond-B} and related, namely QNM frequencies and amplifications factors of acoustic perturbations scattered off ECO-like vortices.

The very definition of QNM requires the amplitude of the ingoing wave at infinity to vanish. It then follows from Eq. \eqref{eq:Rad-INF-large-e} that one must impose
\begin{align}
\label{eq:AN-QNM-condition}
\mathscr{A}+i\mathscr{B}=0.
\end{align}
Eq. \eqref{eq:AN-QNM-condition} can be solved using a simple root-finding algorithm.

On the other hand, according to Eq. \eqref{eq:AmpFactors}, the amplification factors are given by
\begin{equation}
\label{eq:AmpF}
Z_m(\omega,\mathcal{R})=\left|\frac{\mathscr{A}-i\mathscr{B}}{\mathscr{A}+i\mathscr{B}}\right|^2-1.
\end{equation}

\section{Static configurations}
\label{apx:2}
In order to study the static configurations of Eq. \eqref{eq:RadODE-R}, one defines the radial function $L(r)$ \cite{BasakMajumdar:2003}
\begin{equation}
R_{\omega m}(r)=r~e^{\frac{i}{2}\left[(\omega-mB)\log(r^2-1)+2mB\log(r)\right]}L_{\omega m}(r),
\end{equation}
introduces the new coordinate $x=r^2-1$ and sets $\omega=0$, which reduces the radial equation to
\begin{equation}
\label{eq:RadODE-L}
\left[\frac{\D^2}{\D x^2}+\mathcal{W}_1(x)\frac{\D}{\D x}+\mathcal{W}_2(x)\right]L_{\omega m}(x)=0,
\end{equation}
where
\begin{align}
\mathcal{W}_1(x)&=\frac{1}{x}+\frac{1}{x+1},\label{eq:RadODE-L-W1}\\
\mathcal{W}_2(x)&=\frac{1}{4x(x+1)}\left[\frac{m^2B^2}{x}+\frac{1}{x+1}+(1-m^2)\right].\label{eq:RadODE-L-W2}
\end{align}
Eq. \eqref{eq:RadODE-L} is a standard Riemann-Papparitz ODE. Defining
\begin{equation}
L_{\omega m}(x)=x^{\sigma}(x+1)^{-\frac{1}{2}}G_{\omega m}(x),
\end{equation}
with $2\sigma=-imB$, and introducing the new coordinate $y=-x$, one gets
\begin{equation}
\label{eq:RadODE-G}
\left[y(1-y)\frac{\D^2}{\D y^2}+\rho(1-y)\frac{\D}{\D y}-\kappa\lambda\right]G_{\omega m}(y)=0,
\end{equation}
with $2\kappa=-m(1+iB)$, $2\lambda=-m(-1+iB)$ and $\rho=1-imB$. It is easy to check that $\kappa$, $\lambda$ and $\rho$ satisfy the relations $\kappa+\lambda+1=\rho$, $\lambda-\kappa=m$ and $\overline{\rho}=2-\rho$, where $\overline{\rho}$ is the complex conjugate of $\rho$.

The general solution of Eq. \eqref{eq:RadODE-G} reads \cite{HandbookMF}
\begin{align}
G_{\omega m}(y)
=&\mathscr{E}_1{}_2{F}_1(\kappa,\lambda,\rho,y)\nonumber\\&+\mathscr{E}_2y^{1-\rho}{}_2{F}_1(-\lambda,-\kappa,\overline{\rho},y),
\end{align}
where $\mathscr{E}_1,\mathscr{E}_2\in\mathbb{C}$. Equivalently,
\begin{align}
\label{eq:RadR-HOR}
&L_{\omega m}(x)=(x+1)^{-\frac{1}{2}}\cdot\nonumber\\
&[\mathscr{E}_1~x^{\sigma}{}_2{F}_1(\kappa,\lambda,\rho,-x)+\underline{\mathscr{E}_2}~x^{-\sigma}{}_2{F}_1(-\lambda,-\kappa,\overline{\rho},-x)],
\end{align}
where $\underline{\mathscr{E}_2}\equiv(-1)^{1-\rho}\mathscr{E}_2$. The small-$x$ behavior of Eq. \eqref{eq:RadR-HOR} must be written in terms of the radial function $H(r)$ for the BC in Eq. \eqref{QNM-generic-R-Am} to be applied. One can show that 
\begin{align}
H_{\omega m}(r)\sim\mathscr{E}_1e^{+imB r_*}+\underline{\mathscr{E}_2}e^{-imB r_*}
\end{align}
near the would-be acoustic horizon, meaning the BC to be imposed at $r=r_0$ is
\begin{equation}
\label{eq:ABR2}
\underline{\mathscr{E}_2}/\mathscr{E}_1=\mathcal{R}e^{+2imB r_0^*}.
\end{equation}
 
The large-$x$ behavior of Eq. \eqref{eq:RadR-HOR} can be analyzed by making use of the linear transformation formula $15.3.7$ in \cite{HandbookMF} and rearranging the terms. In effect, one gets
\begin{align}
L_{\omega m}(x)=(x+1)^{-\frac{1}{2}}\left[\mathscr{F}_1x^{+\frac{m}{2}}+\mathscr{F}_2x^{-\frac{m}{2}}\right].
\end{align}
with
\begin{align}
\mathscr{F}_1&=\left[\mathscr{E}_1\frac{\Gamma(\rho)}{\lambda\Gamma(\lambda)^2}-\underline{\mathscr{E}_2}\frac{\Gamma(\overline{\rho})}{\kappa\Gamma(-\kappa)^2}\right]\Gamma(\lambda-\kappa),\label{eq:E}\\
\mathscr{F}_2&=\left[\mathscr{E}_1\frac{\Gamma(\rho)}{\kappa\Gamma(\kappa)^2}-\underline{\mathscr{E}_2}\frac{\Gamma(\overline{\rho})}{\lambda\Gamma(-\lambda)^2}\right]\Gamma(\kappa-\lambda).\label{eq:F}
\end{align}
%
It follows from the explicit expression for the energy flux across an arbitrary surface at a constant radial coordinate $r$ that the amplitudes of the ingoing wave and of the outgoing wave are proportional to $(\mathscr{F}_1-i\mathscr{F}_2)$ and $(\mathscr{F}_1+i\mathscr{F}_2)$, respectively \cite{BasakMajumdar:2003}. The absence of ingoing waves at infinity requires that
\begin{equation}
\label{eq:QNM-inf}
\mathscr{F}_1-i\mathscr{F}_2=0,
\end{equation}
which implies that
\begin{align}
\label{coeff-INF}
\frac{\mathscr{E}_1}{\underline{\mathscr{E}_2}}=&\frac{\Gamma(\overline{\rho})}{\Gamma(\rho)}\left[\frac{\Gamma(\kappa)}{\Gamma(-\kappa)}\frac{\Gamma(\lambda)}{\Gamma(-\lambda)}\right]^2\cdot\nonumber\\
&\frac{\lambda\Gamma(-\lambda)^2\Gamma(\lambda-\kappa)-i\kappa\Gamma(-\kappa)^2\Gamma(\kappa-\lambda)}{\kappa\Gamma(\kappa)^2\Gamma(\lambda-\kappa)-i\lambda\Gamma(\lambda)^2\Gamma(\kappa-\lambda)}
\end{align}
On the other hand, Eq. \eqref{QNM-RBC} can be written in terms of the radial function $L_{\omega m}(x)$,
\begin{equation}
\label{eq:QNM-HOR}
\frac{\dot{L}_{\omega m}(x_0)}{L_{\omega m}(x_0)}=-\frac{3}{4(1+x_0)}-\frac{2\cot(\xi)}{\sqrt{1+x_0}}
\end{equation}
where the dot denotes differentiation with respect to $x$. Like Eq. \eqref{eq:QNM-inf}, Eq. \eqref{eq:QNM-HOR} is a condition on the amplitudes $\mathscr{E}_1$ and $\underline{\mathscr{E}_2}$. Explicitly,
\begin{equation}
\label{coeff-r0}
\frac{\mathscr{E}_1}{\underline{\mathscr{E}_2}}=(r_0^2-1)^{-2\sigma}\frac{[(3-4\sigma)r_0^2-3]\tan(\xi)+2r_0(r_0^2-1)}{[(3+4\sigma)r_0^2-3]\tan(\xi)+2r_0(r_0^2-1)}.
\end{equation}
Equating the right-hand side of Eqs. \eqref{coeff-INF} and \eqref{coeff-r0}, one obtains a relation between $r_0$ (or $\delta$), $B$ and $m$. Once $r_0$ (or $\delta$) and $m$ are fixed, one can compute $B_c$, i.e. the critical value of the rotation parameter above which QNMs are unstable. The instability domain is depicted in \autoref{fig:AH-QNM-InsDom}.


\vfill

\bibliography{ECOlike-Vortex}

\end{document}